\documentclass[reprint,aps,pra,longbibliography]{revtex4-2}

\usepackage{graphicx}
\usepackage{natbib}
\usepackage{hyperref}

\usepackage{amsmath,mathtools}
\usepackage{amssymb}

\newcommand{\ket}[1]{\ensuremath{\left| #1 \right\rangle}}

\newcommand{\ii}{\ensuremath{\mathrm{i}}}
\newcommand{\bs}[1]{\mathrm{BS}_{#1}}

\newcommand{\hilb}{\mathcal{H}}

\newcommand{\exch}[1]{\ensuremath{\mathrm{EX}_{#1}}}
\newcommand{\hbs}[2]{\ensuremath{\mathrm{HBS}_{#1, #2}}}

\newcommand{\swapgen}[2]{\ensuremath{\mathrm{SWAP}_{#1, #2}}}

\newcommand{\swap}{\ensuremath{\mathrm{SWAP}}}
\newcommand{\swapt}{\ensuremath{\overline{\mathrm{SWAP}}}}
\newcommand{\sorter}[1]{\ensuremath{\mathrm{S}_{#1}}}

\newcommand{\numbs}[2]{\ensuremath{N^{(#1)}_{\mathrm{BS}}(#2)}}
\newcommand{\numbsA}[3]{\ensuremath{N^{(#1)}_{\mathrm{BS}}(#2, #3)}}

\newcommand{\numdove}[2]{\ensuremath{N^{(#1)}_{\mathrm{Dove}}(#2)}}
\newcommand{\numholo}[2]{\ensuremath{N^{(#1)}_{\mathrm{holo}}(#2)}}
\newcommand{\numphas}[2]{\ensuremath{N^{(#1)}_{\mathrm{phas}}(#2)}}
\newcommand{\nummirr}[2]{\ensuremath{N^{(#1)}_{\mathrm{mirr}}(#2)}}

\usepackage[utf8]{inputenc}

\newcommand{\fftp}[1]{\ensuremath{F_{\mathrm{path}}^{(#1)}}}
\newcommand{\fftpn}{\ensuremath{F_{\mathrm{path}}}}
\newcommand{\fp}{\ensuremath{F_{P}}}
\newcommand{\ffto}[1]{\ensuremath{F_{\mathrm{OAM}}^{(#1)}}}

\newcommand{\dmin}{\ensuremath{d_{\mathrm{min}}}}
\newcommand{\dmax}{\ensuremath{d_{\mathrm{max}}}}
\newcommand{\din}{\ensuremath{d_{\mathrm{in}}}}
\newcommand{\dout}{\ensuremath{d_{\mathrm{out}}}}

\newcommand{\mults}{\ensuremath{\mu_{\mathrm{S}}}}
\newcommand{\multswap}{\ensuremath{\mu_{\swap{}{}}}}

\usepackage{color}
\newcommand{\red}[1]{{#1}}

\usepackage{ulem} % for \sout command

\renewcommand{\emph}[1]{\textit{#1}}

\begin{document}

\title{High-dimensional quantum Fourier transform of twisted light}
\author{Jaroslav Kysela}
\affiliation{University of Vienna, Faculty of Physics, Boltzmanngasse 5, Vienna, Austria}
% see email: "Reminder: University of Vienna Affiliation Policy" from 28. 1. 2021
\affiliation{Institute for Quantum Optics and Quantum Information, Austrian Academy of Sciences, Boltzmanngasse 3, Vienna, Austria}
\date{\today}
\begin{abstract}
    The Fourier transform proves indispensable in the processing of classical information as well as in the quantum domain, where it finds many applications ranging from state reconstruction to prime factorization. An implementation scheme of the $d$-dimensional Fourier transform acting on single photons is known that uses the path encoding and requires $O(d \log d)$ optical elements. In this paper we present an alternative design that uses the orbital angular momentum as a carrier of information and needs only $O(\sqrt{d}\log d)$ elements, rendering the path-encoded design inefficient. The advantageous scaling and the fact that our approach uses only conventional optical elements allows for the implementation of a 256-dimensional Fourier transform with the existing technology. Improvements of our design, as well as explicit setups for low dimensions, are also presented.
\end{abstract}

\maketitle

\section{Intro}

The Fourier transform is arguably one of the most important tools in modern mathematics, science and engineering. Its applications range from a purely mathematical use in differential calculus \cite{osgood2019lectures} to modelling optical properties of light such as a free-space propagation or a propagation through a system of lenses \cite{tysonFourier}. On a more practical level, the Fourier transform is used heavily in the spectral analysis of audio and video signals \cite{signalSystems}. %as it links the frequency and time domains. 
Its discrete version lies at the heart of various data-processing algorithms used by modern computers. Popular image and audio compression algorithms JPEG and MP3 \cite{jpeg,mp3} are notable examples. The Fourier transform plays also a central role in the quantum domain. Not only it underlies the duality of the position and momentum representations in quantum physics \cite{dirac_principles_2009}, but it also proves useful in the quantum information processing. The Shor's factoring algorithm, based on the quantum Fourier transform, dramatically outperforms any of its classical counterparts \cite{Shor1994}. The quantum Fourier transform is often understood as a transformation applied to a many-particle quantum state. There are, however, many cases when the transform has to act on a large state of a single particle. Such a single-particle quantum Fourier transform finds a plethora of applications, such as the generation of mutually unbiased bases in quantum state tomography and quantum key distribution \cite{Wootters1989, brierley2010, DURT2010, Giovannini2013, Groblacher2006, Malik2012}, generation of angular states \cite{Barnett1990, Franke-Arnold2004, Yao2006, Wang2017}, and implementation of programmable universal multiport arrays \cite{lpezpastor2019arbitrary, Saygin2020, pereira2020universal}.

Various systems can be chosen as quantum carriers of information. Such a system can be, for example, a single photon, where the information is encoded into its orbital angular momentum (OAM). The OAM of a photon is manifested by a helical structure of the wavefront of the photon's wavefunction \cite{Allen1992, Krenn2014, Erhard2018review}, for which it is sometimes dubbed `twisted light.' Each twist of the helix corresponds to a quantum of OAM. The number of these quanta is not bounded. Unlike in the classical computation, which uses only two values---0 and 1---to encode the data, the OAM allows one to define a $d$-valued alphabet for arbitrary finite integer $d$ \cite{Erhard2018review}. The data is thus represented as a quantum superposition of $d$ different OAM eigenstates of a single photon. Many experiments have been conducted in various contexts \cite{Lavery2012, OSullivan2012, Malik2012, Mirhosseini2013, Wang2017, Brandt2020}, where the single-photon OAM Fourier transform is implemented using specially designed optical elements with nontrivial phase profiles \cite{Berkhout2010}. Recently, an alternative approach has been found that relies on interferometers and requires only conventional optical elements familiar from the classical optics \cite{oamfft}. This approach is recursive and allows for an efficient implementation when the dimension of the OAM space is of the form $d = 2^M$ for some $M \in \mathbb{N}$.

In this paper, we present an improved interferometric design of the Fourier transform inspired by that of Ref.~\cite{oamfft}, but contrary to the original proposal, our design avoids recursion. As a result, we not only simplify the resulting scheme, but also determine analytically the optimal decomposition and, most importantly, reduce dramatically the number of optical elements in the setup. The scheme of Ref.~\cite{oamfft} requires asymptotically $6.1412 \, d$ beam splitters, whereas we present a scheme that requires only $(11/4) \sqrt{d}\log(d)$ of them and allows for further simplifications. Even though the following discussion focuses on the quantum regime, the identical physical setups can be also used to implement the Fourier transform of the OAM states of classical light.

This manuscript is organised as follows. After the brief summary of the mathematical formalism behind our scheme in section \ref{sec:fft}, we present the setup of the OAM Fourier transform in section \ref{sec:implementation}, \red{whose scaling is studied in section \ref{sec:scaling}}. In section \ref{sec:higherspaces} we discuss the periodicity of our implementation. In section \ref{sec:polenhanced} an improved version of our scheme is presented, which uses polarization. Section \ref{sec:oamenhanced} shows how to modify our setup to implement the path-only Fourier transform and in section \ref{sec:conclusion} we summarize our results.

\section{Fourier transform of OAM}
\label{sec:fft}

\begin{figure}
    \centering
    \includegraphics[width=\linewidth]{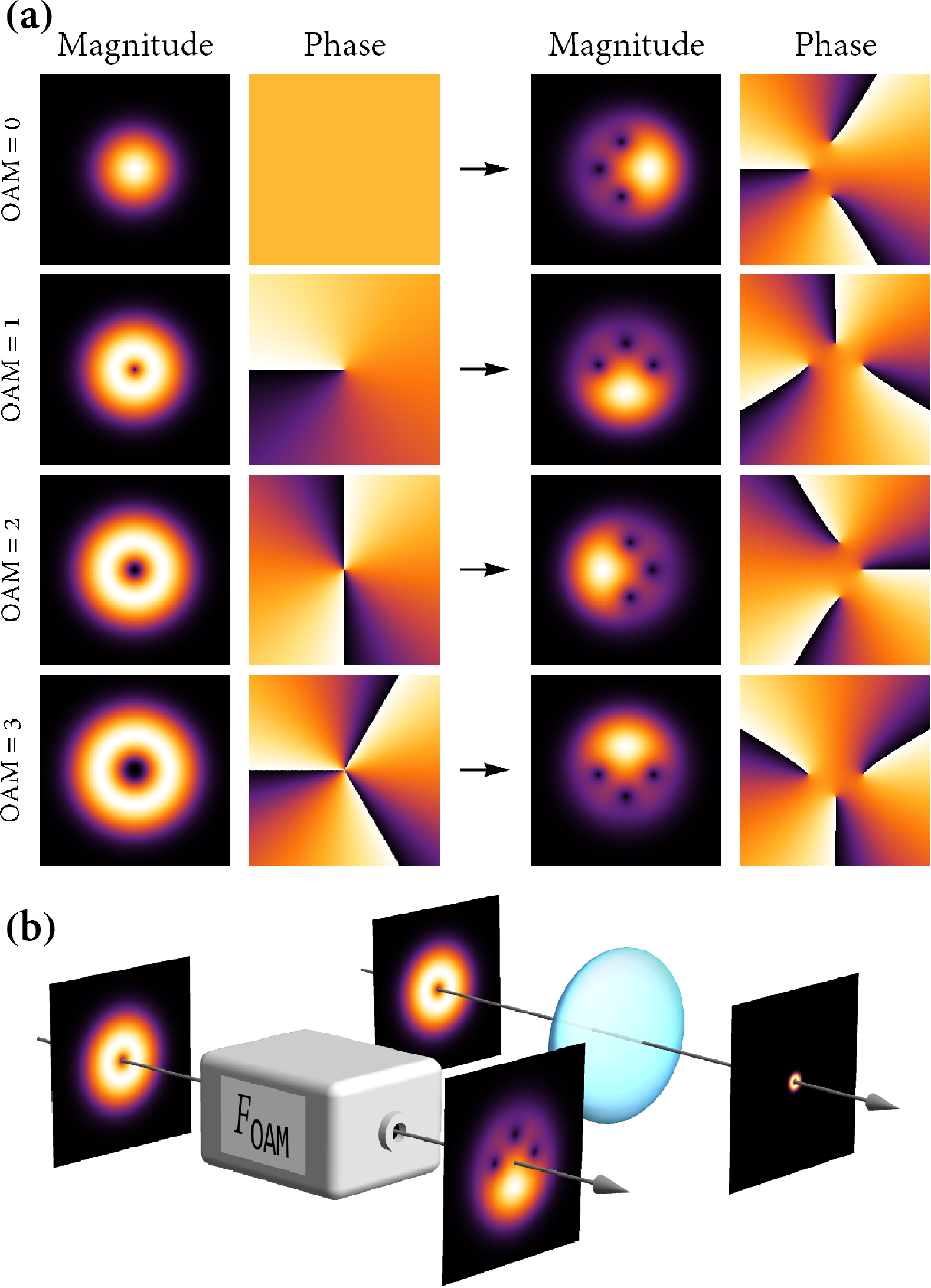}
    \caption{Fourier transform of the orbital angular momentum of light. (a) For illustration, explicit forms of the incoming and resulting states are shown, when the four-dimensional Fourier transform is applied to OAM eigenstates $\ket{0}$, $\ket{1}$, $\ket{2}$, and $\ket{3}$. The first and third columns depict magnitudes, whereas the second and fourth columns depict phases of the light beam. (b) The OAM Fourier transform should not be mistaken for the Fourier property of a lens. Resulting spatial profiles of a light beam are drastically different.}
    \label{fig:fft_of_oam}
\end{figure}

\begin{figure*}
    \centering
    \includegraphics[width=\linewidth]{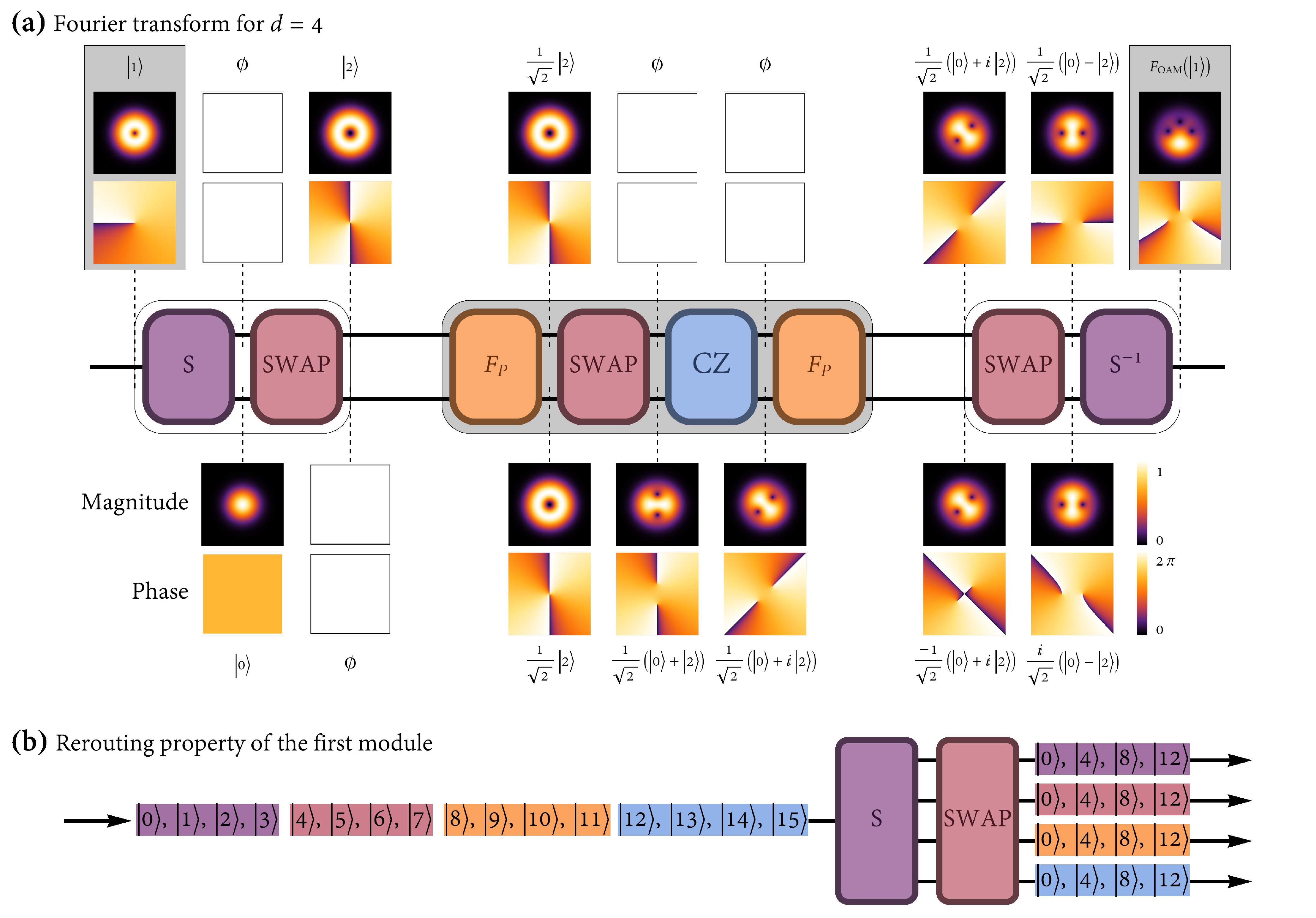}
    \caption{The setup of the OAM Fourier transform. (a) The evolution of the incoming OAM eigenstate $\ket{1}$ when subjected to the 4-dimensional OAM Fourier transform with $d_A = d_B = 2$. The final state is equal to $F_{\mathrm{OAM}} (\ket{1}) = (\ket{0}-i \ket{1}-\ket{2}+i \ket{3})/2$. The insets depict magnitudes and phases of the spatial transversal profiles of the light beam propagating along upper and lower paths. The associated kets are also shown. When there is a blank inset, there is no light at that particular stage of the propagation. The multiplicities of eigenstates are already taken into account, see section \ref{sec:implementation} (b) The first module of the setup, consisting of an OAM sorter and a swap operator, reroutes incoming OAM eigenstates into multiple paths, such that OAM eigenstates $\ket{0}, \ldots, \ket{d_B - 1}$ leave the swap via the first path, eigenstates $\ket{d_B}, \ldots, \ket{2 d_B - 1}$ leave via the second path etc. Note also that the module changes the number of OAM quanta for each incoming eigenstate. Here, an explicit case for $d_B = 4$ is shown. Even though the multiplicity of incoming OAM eigenstates is 1, the multiplicity of eigenstates leaving the module is equal to 4.}
    \label{fig:evolution}
\end{figure*}

Each OAM eigenstate is characterized by a specific spatial profile of the photon's wavefunction, whose phase has a helical structure with a singularity in the center, whereas the amplitude resembles a doughnut \cite{Allen1992,Erhard2018review}, see Fig.~\ref{fig:fft_of_oam}. We consider a single photon that carries information encoded into a quantum superposition of $d$ different OAM eigenstates $\ket{0}$, \ldots, $\ket{d-1}$. Such eigenstates span a $d$-dimensional Hilbert space $\hilb$. To characterize how the Fourier transform acts on a general superposition of eigenstates, it suffices to specify its action on the basis states $\ket{k}$ for $k = 0, \ldots, d-1$. The $d$-dimensional Fourier transform of an OAM eigenstate $\ket{k}$ is defined as 
\begin{equation}
    \ffto{d}(\ket{k}) = \frac{1}{\sqrt{d}} \sum_{j=0}^{d-1} e^{i \, 2 \pi j k / d} \ket{j}. \label{eq:definition}
\end{equation}
The transform turns the helical profile of an eigenstate into a complicated wavefront with many singularities as demonstrated in Fig.~\ref{fig:fft_of_oam}(a). One should not mistake the Fourier transform of spatial modes of light for the Fourier property of a normal lens, see Fig.~\ref{fig:fft_of_oam}(b). For example, the four-dimensional Fourier transform applied to the OAM eigenstate $\ket{1}$ results in the state $(\ket{0} + i \ket{1} - \ket{2} - i \ket{3})/2$ with a non-trivial wavefront. On the contrary, the OAM eigenstates, including $\ket{1}$, preserve their overall shape when propagating through a lens.

To implement the transform in Eq.~\eqref{eq:definition}, we take advantage of its algebraic structure. It is easy to show that the discrete $d$-dimensional Fourier transform can be expressed as a composition of two lower-dimensional Fourier transforms \cite{Cooley1965}. One can apply this decomposition recursively such that in the end the Fourier transform is decomposed into elementary 2-dimensional blocks. In the physical scenario, these elementary blocks can be identified with beam splitters. As a result, the $d$-dimensional Fourier transform acting on path-encoded qudits can be efficiently implemented \cite{Torma1996}. Similar ideas can be applied when implementing the $d$-dimensional Fourier transform in the OAM, see Fig.~1 in Ref.~\cite{oamfft}. In that case, the initial superposition of OAM eigenstates of an incoming photon is first transformed into a superposition of a smaller number of OAM eigenstates propagating along multiple paths. This way, the information stored in the state gets redistributed between the path and OAM degrees of freedom. The Fourier transform is then implemented by an application of a lower-dimensional Fourier transform that acts only on the OAM degree of freedom, followed by another lower-dimensional Fourier transform that acts only on the path degree of freedom. The whole procedure is finished by recombining all the resulting OAM eigenstates into a single output path. 

To implement the lower-dimensional OAM Fourier transform, the same decomposition is applied recursively until a series of elementary 2-dimensional blocks is reached. Such a scheme for the OAM Fourier transform has been shown recently to provide savings in resources when compared to alternative approaches \cite{oamfft}. In this paper, we use the same decomposition of the Fourier transform as in Ref.~\cite{oamfft}, but unlike in their approach, we do not apply this decomposition recursively.

The decomposition relies on a factorization of the initial $d$-dimensional Hilbert space $\hilb$ of the OAM of light into two subspaces $\hilb_O$ and $\hilb_P$, such that $\hilb = \hilb_O \otimes \hilb_P$. The state $\ket{k}$ in $\hilb$ is then identified with states $\ket{q}_O \in \hilb_O$ and $\ket{r}_P \in \hilb_P$ as follows
\begin{eqnarray}
        \ket{k} \in \hilb \quad \longleftrightarrow \quad \ket{q}_O \otimes \ket{r}_P \in \hilb_O \otimes \hilb_P. \label{eq:factorization}
\end{eqnarray} 
Let us denote the dimensions of $\hilb_O$ and $\hilb_P$ by $d_A$ and $d_B$, respectively, such that $d = d_A \, d_B$. The Fourier transform acting on a $d$-dimensional Hilbert space $\hilb$ can be expressed as a product of four operators that are applied from right to left
\begin{equation}
    \ffto{d} = \fftp{d_A} \cdot \mathrm{CZ} \cdot \swapt \cdot \fftp{d_B}. \label{eq:main_eq}
\end{equation}
At first, a $d_B$-dimensional path-only Fourier transform is applied only to the subspace $\hilb_P$. The physical implementation of the path-only Fourier transform is henceforth denoted by $\fp$. Then, a swap operator with $d_B$ input paths and $d_A$ output paths exchanges states of the two subspaces. Its action on an input mode reads $\swapt(\ket{r}_O\ket{q}_P) = \ket{q}_O\ket{r}_P$. The overline is used to distinguish the mathematical operation $\swapt$ from its physical implementation, which is discussed later on and which is denoted by $\swap$ in the following. The swap operator effectively refactorizes the Hilbert state from $\hilb_O \otimes \hilb_P$ into $\hilb'_O \otimes \hilb'_P$, where now $\dim \hilb'_O = d_B$ and $\dim \hilb'_P = d_A$. In the third step, a high-dimensional controlled-Z gate $\textrm{CZ}$ is applied, where the path plays the role of the $d_A$-dimensional control qudit and the OAM degree of freedom represents the $d_B$-dimensional target qudit 
\begin{equation}
    \textrm{CZ}(\ket{m}_O\ket{l}_P) = (Z^{l}\ket{m}_O)\ket{l}_P = \omega^{m \, l} \ket{m}_O \ket{l}_P,
    \label{eq:formula_phase_gate}
\end{equation}
where $\omega = \exp{(2 \pi \ii/d)}$. The sequence of operators is concluded by the second, this time $d_A$-dimensional, path-only Fourier transform. For the proof of Eq.~\eqref{eq:main_eq} see Appendix \ref{sec:derivation}. To illustrate the action of individual components in Eq.~\eqref{eq:main_eq} for incoming OAM eigenstates, the propagation of the OAM eigenstate $\ket{1}$ through the setup of $4$-dimensional Fourier transform is demonstrated in Fig.~\ref{fig:evolution}(a).

The decomposition on the righ-hand side of Eq.~\eqref{eq:main_eq} assumes that the initial Hilbert space is already factorized into $\hilb_O$ and $\hilb_P$, where $\hilb_O$ is spanned by $d_A$ different OAM eigenstates and $\hilb_P$ is spanned by $d_B$ different propagation modes. In the actual setup this factorization is performed by an additional module, which consists of a $d_A$-dimensional OAM sorter $\sorter{}$ \cite{oamfft} and a swap operator with $d_A$ input and $d_B$ output paths. The operation of this module is graphically illustrated in Fig.~\ref{fig:evolution}(b). Analogously, the final recombination is taken care of by another module that has the inverse structure to that of the first module. It comprises a swap with $d_A$ input and $d_B$ output paths and a $d_B$-dimensional OAM sorter $\sorter{}^{-1}$ that is operated in reverse. To emphasize the role of the center module whose action is given by Eq.~\eqref{eq:main_eq}, we will refer to it as the Fourier transform \emph{proper}. All three modules---the factorization module, the Fourier transform proper, and the recombination module---are shown explicitly in Fig.~\ref{fig:evolution}(a). For details see the next section and Appendix \ref{sec:explicit_setups}.

As in Ref.~\cite{oamfft}, we consider throughout the paper only dimensions $d$ that are powers of two, i.e., $d = 2^M$ for some integer $M$. It remains an open question still, how to implement efficiently the Fourier transform for a general dimension. 

\section{Improved scheme}
\label{sec:implementation}

\begin{figure*}
    \centering
    \includegraphics[width=\linewidth]{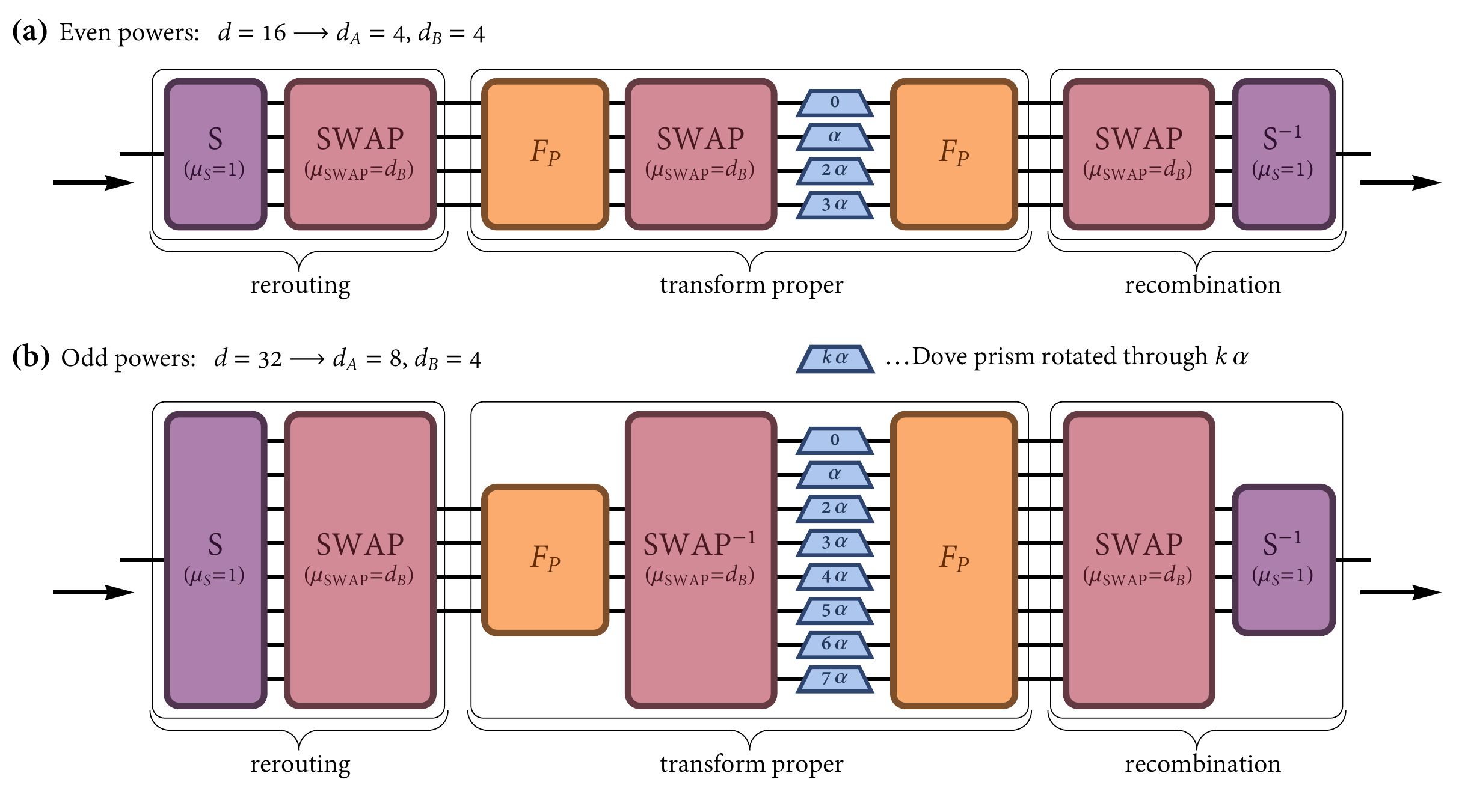}
    \caption{The two classes of setups for the Fourier transform of the OAM of light. The structure of the setup for dimensions of the form $d = 2^M$ with $M \in \mathbb{N}$ follows from Eq.~\eqref{eq:main_eq} and depends on the parity of $M$. (a) For even $M$ one can take $d_A = d_B = \sqrt{d}$. The whole setup has then a fixed number $\sqrt{d}$ of paths and can be divided into three modules. The OAM eigenstates entering the setup from the left are initially rerouted into different paths in the first module. The second module performs the Fourier transform proper and the third module recombines all the OAM eigenstates into a single output path. The figure is shown for a specific example of $d = 16$, for which $d_A = d_B = 4$. The Dove prisms are rotated through the angle $l \alpha = l \, \pi/(d_A \, d)$ and are supplemented with mirrors, which are not shown. (b) The setup for odd $M$ is analogous to that in (a) with the only difference that we take $d_A = \sqrt{2d}$ and $d_B = \sqrt{d/2} = d_A/2$. The number of paths therefore changes throughout the setup. In the figure a specific example for $d = 32$ is shown, for which $d_A = 8$ and $d_B = 4$. For details see the main text.}
    \label{fig:fft_paral_even_odd}
\end{figure*}

\begin{figure}
    \centering
    \includegraphics[width=\linewidth]{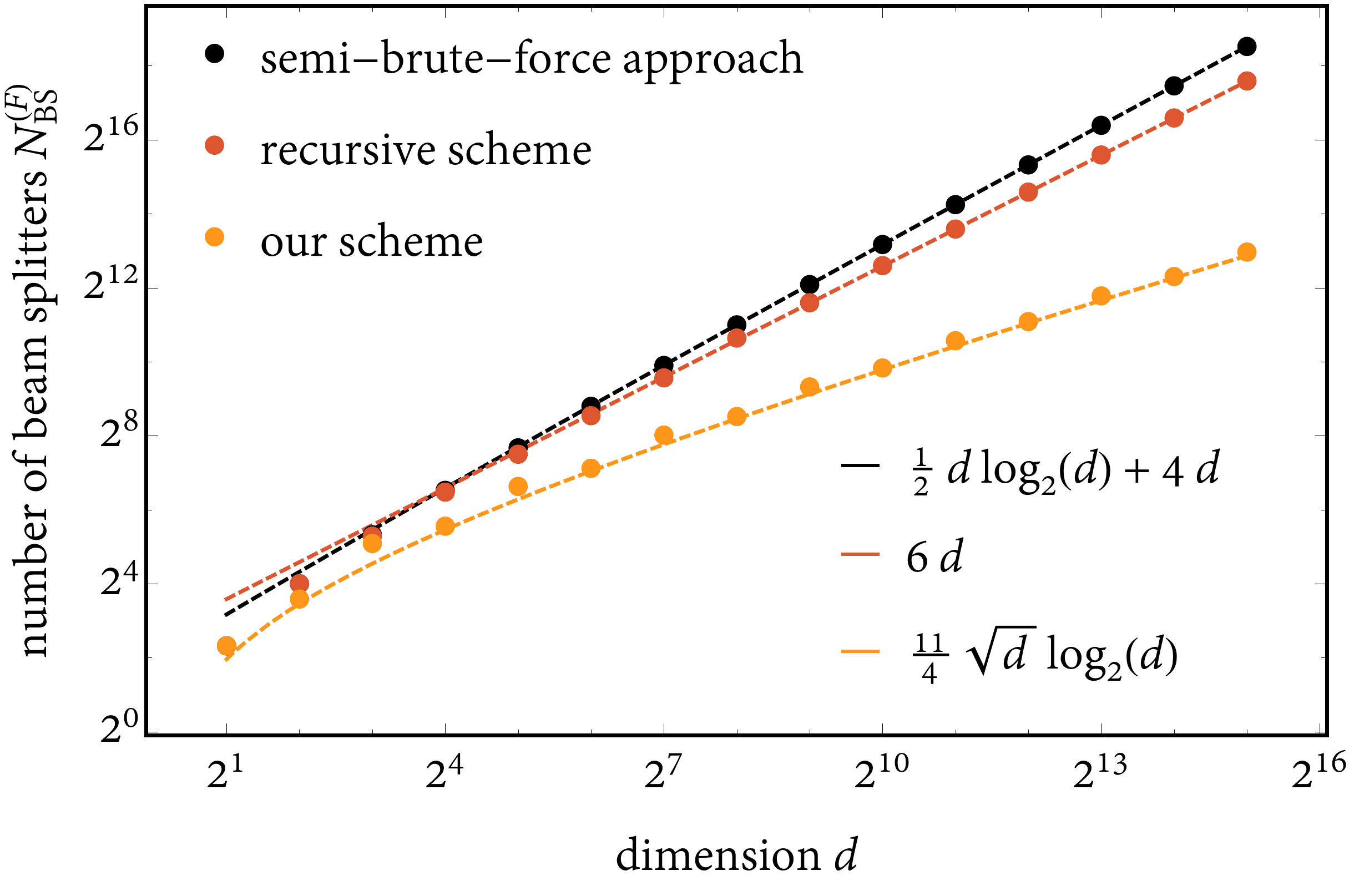}
    \caption{The number of beam splitters in the setup for the OAM Fourier transform as a function of the dimension $d$, using three different approaches. The semi-brute-force approach requires $O(d \log_2(d))$ beam splitters. The recursive scheme of Ref.~\cite{oamfft} requires only $O(d)$ beam splitters. Nevertheless, as is evident from the plot, our scheme is considerably more efficient than both the semi-brute-force and recursive schemes as it scales as $O(\sqrt{d}\log_2(d))$. Note that both axes in the plot use logarithmic scaling. The point markers denote precise numbers, whereas lines represent analytically the growing trends.}
    \label{fig:comparison_bs}
\end{figure}

Our scheme requires only conventional optical elements for its construction. Namely, mirrors, beam splitters, Dove prisms, phase shifters, and holograms \red{\footnote{\red{Throughout this paper, by a hologram we mean an element that adds or subtracts a finite (integer) number of OAM quanta and has thus a relatively simple helix phase profile. Such an element can be a spiral phase plate \cite{BEIJERSBERGEN1994321,Oemrawsingh04}  %\cite{BEIJERSBERGEN1994321,TURNBULL1996183,Oemrawsingh04}
or a fixed fork diagram \cite{heckenberg_laser_1992}. Importantly, no complex phase profiles are necessary that have to be projected on a spatial light modulator (SLM), even though SLMs could also be used in our implementation, of course.}}}. In this section we demonstrate how to use these optical elements to implement individual components of the setup. As explained in the previous section and illustrated in Fig.~\ref{fig:evolution}(a), the setup of the OAM Fourier transform can be divided into three modules. The first module reroutes the incoming OAM eigenstates $\ket{0}, \ldots, \ket{d - 1}$ into $d_B$ different paths. Its role is to factorize the initial Hilbert space into two subspaces $\hilb_O$ and $\hilb_P$ as described in \eqref{eq:factorization}. It turns out, see Appendix \ref{sec:optimal_choice}, that the roles of $d_A$ and $d_B$ are symmetric and we can without loss of generality assume that $d_A \geq d_B$ from now on. This simplifies the following discussion. The first module consists of a $d_A$-dimensional OAM sorter and a swap operator that has $d_A$ input and $d_B$ output ports. 
    
The $d_A$-dimensional sorter sorts individual incoming OAM eigenstates $\ket{0}_O, \ldots, \ket{d_A - 1}_O$ into different output ports $\ket{0}_P, \ldots, \ket{d_A - 1}_P$. Multiple designs of the OAM sorter have been investigated in experiments using different physical principles, such as multi-plane light conversion \cite{Fontaine2019,Labroille2014}, light scattering in random media \cite{Fickler2017} or light propagation through two phase plates with specially designed wavefronts \cite{Berkhout2010}. In this paper we employ the interferometric design of Leach \textit{et al.} \cite{Leach2002, Leach2004, Garcia-Escartin2008, Erhard2017, Xgate}. Intrinsic to this design is its modulo property \cite{oamfft}, where also eigenstates with more than $d_A$ quanta of OAM get sorted into one of $d_A$ output ports. This way, the sorter separates the $d$ incoming OAM eigenstates into $d_A$ groups, each of which propagates along a different path and contains $d_B$ OAM eigenstates. The side effect of the modulo property is that the OAM eigenstates leaving the sorter are of the form $\ket{0}, \ket{d_A}, \ket{2 d_A}, \ldots, \ket{(d_B - 1) \, d_A}$. In other words, the difference between two successive values of OAM is equal to $d_A$. We call this difference a \emph{multiplicity} of the OAM eigenstates. The internal structure of individual components of the setup, such as swaps, has to be adjusted to this multiplicity. Suppose that the multiplicity of eigenstates entering the sorter is $\mu \in \mathbb{N}$. In order for the sorter to sort these states correctly, the order of each OAM exchanger \cite{oamfft} in the OAM sorter has to be multiplied by $\mu$. Since the number $\mu$ affects the construction of the sorter, we call it the multiplicity of the sorter and denote it by $\mults$. The multiplicity of the eigenstates leaving the sorter is then $\mults \, d_A$. Nevertheless, in our scheme we assume that the multiplicity of incoming OAM eigenstates is equal to one and the default sorter structure can be used.

The second part of the first module consists of a swap operator with $d_A$ input and $d_B$ output ports. Its interferometric implementation, denoted by $\swap$, is given in Ref.~\cite{oamfft} and is briefly summarized in Appendix \ref{sec:explicit_setups}. When there is the same number of input and output ports, i.e., $d_A = d_B$, both forward and backward passes through $\swap$ represent the same operation, see also Fig.~\ref{fig:fft_paral_even_odd}(a). The multiplicity of OAM eigenstates leaving the swap stays $\multswap = d_A = d_B$ in such a case. When $d_A > d_B$, the self-inverse property of $\swap$ is lost. The multiplicity of incoming OAM eigenstates is $d_A$ and the implementation acts like
\begin{equation}
    \swap\left(\ket{d_A \cdot l}_O\ket{m}_P \right) = \ket{d_B \cdot m}_O\ket{l}_P.
\end{equation}
The multiplicity of the output OAM eigenstates thus becomes $d_B$. From the construction of $\swap$ it follows that we have to set its multiplicity to the multiplicity of the output eigenstates, $\multswap = d_B$. To summarize the action of the first module, we obtain transformation rules
\begin{eqnarray}
    \ket{q} \in \hilb \quad \longrightarrow \quad \ket{d_B \cdot m}_O \otimes \ket{l}_P \in \hilb_O \otimes \hilb_P,
    \label{eq:tensor_index}
\end{eqnarray}
where $q = d_A \, l + m$ is the initial number of OAM quanta of the incoming OAM eigenstate, $0 \leq m < d_A$, and $0 \leq l < d_B$. For details see Appendix \ref{sec:sorter_swap}. 

The second module represents the Fourier transform proper and comprises four operations as given by Eq.~\eqref{eq:main_eq}. These operations are implemented as follows:

\begin{enumerate}
    \item The $d_B$-dimensional path-only Fourier transform is represented in the setup by block $\fp$ with $d_B$ input and output paths. Its efficient implementation in terms of beam splitters and phase shifters is presented in Ref.~\cite{Torma1996}. Additional mirrors are necessary in our implementation of $\fp$ as explained in Ref.~\cite{oamfft}. Block $\fp$ acts on propagation modes and leaves OAM eigenstates unaffected.
    
    \item The swap operator is implemented analogously to the swap in the first module. It has $d_B$ input and $d_A$ output ports. For reasons that are explained in Appendix \ref{sec:sorter_swap} it is advantageous to assume that the number of output ports of $\swap$ never exceeds the number of its input ports. This poses a problem in the case when $d_A > d_B$. The easy fix is to use the implementation backward. Instead of $\swap$ we thus use $\swap^{-1}$, see Fig.~\ref{fig:fft_paral_even_odd}(b).
    
    \item The controlled-Z gate CZ is implemented as a series of properly rotated Dove prisms, each of which lies on a different path. Specifically, a Dove prism in the $k$-th path is rotated through angle $k \alpha$, where $\alpha = \pi/(d_A \, d)$. As a Dove prism also reverses the sign of the OAM value, an additional mirror supplements each Dove prism to revert the sign back.
    
    \item The $d_A$-dimensional path-only Fourier transform is represented by block $\fp$ with $d_A$ input and output paths. It is constructed using the same design as the initial $d_B$-dimensional path-only Fourier transform.
\end{enumerate}
This module is reminiscent of the schemes presented in Refs.~\cite{PhysRevA.71.042324,Ionicioiu2016} where the OAM equivalent of the polarizing beam splitter was studied.

The third module recombines all resulting OAM eigenstates into a single output path. It has the inverse structure to that of the first module. It consists of a swap operator with $d_A$ input paths, $d_B$ output ports, and multiplicity $\multswap = d_B$, followed by a $d_B$-dimensional OAM sorter that is operated in reverse and that has multiplicity $\mults = 1$. This setup implements relations
\begin{eqnarray}
     \ket{d_A \cdot j}_O \otimes \ket{k}_P \in \hilb_O \otimes \hilb_P \quad \longrightarrow \quad \ket{r} \in \hilb,
    \label{eq:tensor_index_rev}
\end{eqnarray}
where $r = d_B \, k + j$ is the final number of OAM quanta of a particular OAM eigenstate, $0 \leq k < d_A$, and $0 \leq j < d_B$.

An important observation is that values of $d_A$ and $d_B$ are not a priori fixed. The only condition these dimensions have to satisfy is $d = d_A \ d_B$. This gives us freedom to choose such values that lead to the minimal number of optical elements in the physical implementation. In Appendix \ref{sec:optimal_choice} an analytical proof is presented, where we show that the optimal choice of $d_A$ and $d_B$ depends on the parity of $M$ in $d = 2^M$. For an even power $M$ one obtains
\begin{equation}
  \text{even powers}: \quad d_A = d_B = \sqrt{d}.
  \label{eq:even}
\end{equation}
The number of paths in the whole setup implementing the OAM Fourier transform is then constant and equal to $\sqrt{d}$. A specific example of such a setup for $M = 4$ is depicted in Fig.~\ref{fig:fft_paral_even_odd}(a). For odd $M$ the optimal values are
\begin{equation}
  \text{odd powers}: \quad d_A = \sqrt{2 d}, \quad d_B = \sqrt{d/2}.
  \label{eq:odd}
\end{equation}
In this case, the number of paths in the setup alternates between $d_A$ and $d_B = d_A/2$. A specific example of such a setup for $M = 5$ is depicted in Fig.~\ref{fig:fft_paral_even_odd}(b).

\section{Scaling and losses}
\label{sec:scaling}

The setups with optimal values of $d_A$ and $d_B$\red{, see Eqs.~\eqref{eq:even} and \eqref{eq:odd},} have the minimal number of beam splitters, Dove prisms, holograms, and phase shifters among all setups that implement the $d$-dimensional Fourier transform using decomposition \eqref{eq:main_eq}. In accord with Ref.~\cite{oamfft} we focus on the number of beam splitters as these affect significantly the interferometric stability of the final setup. In our scheme, this number is approximately equal to
\begin{eqnarray}
    \numbs{F}{d} \approx \frac{11}{4} \sqrt{d}\log(d).
\end{eqnarray}
For other types of optical elements we obtain the same scaling $O(\sqrt{d}\log_2(d))$, as demonstrated in detail in Appendix \ref{sec:other_elements}. There exist at least two alternative approaches based on interferometers that can be used to construct the OAM Fourier transform. The brute-force approach consists in transforming all incoming OAM eigenstates into the path encoding and then applying a path-only Fourier transform. For dimensions of the form $d = 2^M$, which we consider in this paper, an efficient design of Ref.~\cite{Torma1996} can be used to implement the path-only Fourier transform. Such a semi-brute-force approach requires asymptotically $O(d \log_2(d))$ beam splitters. The other approach of Ref.~\cite{oamfft}, relying on the recursive construction of the OAM Fourier transform, requires $O(d)$ beam splitters. The comparison of our approach and the two alternatives is presented in Fig.~\ref{fig:comparison_bs}. \red{Recently, a quantum version of the integer fast Fourier transform has been presented \cite{asaka_quantum_2020}, which needs $O(d \log_2 (d))$ gates, but works in parallel on a large number of data sets with no additional overhead. We compare this scheme with the parallelization of our scheme in the next section.}

\red{The practical applicability of any optical design is limited by losses of the actual implementation. The detailed analysis of losses of our design lies beyond the scope of the present paper. Nevertheless, we provide some rough estimates of the setup's performance.} 
As pointed out in Ref.~\cite{ClementsScheme} for general unitaries, the losses suffered by photons propagating through a network of interferometers can be significantly reduced when the topology of the network is rectangular, as opposed to the triangular topology \cite{Reck1994}. The losses increase with the number of beam splitters a photon has to traverse. This number is quantified by the depth of the network \cite{ClementsScheme}, which scales like $d$ and $2d$ for \cite{ClementsScheme} and \cite{Reck1994}, respectively. In our implementation of the Fourier transform, the setup has a rectangular structure whose depth scales like \red{$(15/2) \log_2(d)$, i.e. $(15/2) M$, for both even and odd $M$.} 
\red{In Appendix \ref{sec:losses}, a simple model is presented, with the help of which we assess the approximate effect of losses in our design.}

There are at most $d_A \sim \sqrt{d}$ paths \red{in the setup}, which is especially evident for dimensions $d = 2^M$ with even exponent $M$, see Fig.~\ref{fig:fft_paral_even_odd}. The reduced number of propagation modes is another improvement over alternative approaches. The semi-brute force has also a rectangular structure, but requires $d$ paths. Similarly, the setup based on the recursive scheme \cite{oamfft} begins with an OAM sorter that reroutes all incoming OAM eigenstates into $d$ different paths. The rest of the scheme has a triangular structure with a single final output path.

\section{Higher-OAM subspaces}
\label{sec:higherspaces}

\red{A distinctive feature of our implementation is that it works not only for OAM eigenstates it was designed for (and their superpositions), but also for eigenstates that lie outside of the original range of OAM values. The scheme acts primarily on states lying in a subspace spanned by OAM eigenstates $\ket{0}, \ket{1}, \ket{2} \ldots, \ket{d-1}$. Any superposition of these basis states is thus transformed in compliance with Eq.~\eqref{eq:definition}. Consider a new subspace spanned instead by OAM eigenstates of the form $\ket{0 + a \, d}, \ket{1 + a \, d}, \ket{2 + a \, d} \ldots, \ket{(d-1) + a \, d}$, where $a$ is a fixed integer. For example, when $a = 1$, the new subspace contains all possible superpositions of eigenstates $\ket{d}, \ket{d+1}, \ket{d+2} \ldots, \ket{2d-1}$.} It is straightforward to prove that the setup for the $d$-dimensional OAM Fourier transform acts on such OAM eigenstates like
\begin{equation}
    \ffto{d}(\ket{k + a \, d}) = e^{\frac{2\pi i}{d_A} \, m \, a} \cdot \frac{1}{\sqrt{d}} \sum_{j=0}^{d-1} e^{i \, 2 \pi j k / d} \ket{j + a \, d}, \label{eq:fft_higher}
\end{equation}
where $m = k \mod d_A$. From the formula above it follows that the setup applies a $d$-dimensional Fourier transform to the new subspace in a way that is completely analogous to the way it acts on the original subspace. The only difference is an extra phase factor $e^{2 \pi i m a/d_A}$, which emerges when a higher-OAM eigenstate propagates through Dove prisms in the second module of the setup. Whenever $a$ is a multiple of $d_A$, the phase factor vanishes and one obtains the exact $d$-dimensional Fourier transform.

The periodicity of our setup can be employed to apply the Fourier transform simultaneously to many OAM subspaces. Each value of parameter $a$ defines a particular $d$-dimensional subspace of OAM eigenstates. This value is for one specific subspace fixed, but can be otherwise chosen arbitrarily \red{\footnote{The only limitation is given by physical considerations, since the spatial extent of the OAM eigenstate increases with the number of OAM quanta and for large numbers is the manipulation of the photon impractical \cite{Campbell_12,Shen_2013,Fickler13642}.}}. As a consequence, our setup applies the $d$-dimensional Fourier transform simultaneously to potentially enormous number of subspaces. One can imagine to have data stored in several data sets that are represented as one large superposition of OAM eigenstates of a single photon. It is then sufficient to send the photon only once through the setup in order to apply the Fourier transform separately and simultaneously to each data set. \red{To illustrate this point, consider two data sets, $\{\alpha_0, \ldots, \alpha_3\}$ and $\{\beta_0, \ldots, \beta_3\}$, each consisting of four complex numbers. These data sets can be (after proper normalization) encoded into the following state of a single photon
\begin{multline}
    \ket{\psi} = \alpha_0 \ket{0} + \alpha_1 \ket{1} + \alpha_2 \ket{2} + \alpha_3 \ket{3} \\ + \beta_0 \ket{8} + \beta_1 \ket{9} + \beta_2 \ket{10} + \beta_3 \ket{11},
    \label{eq:amplenc}
\end{multline}
where eigenstates $\ket{0}, \ldots, \ket{3}$ span the original 4-dimensional subspace and eigenstates $\ket{8}, \ldots, \ket{11}$ span another 4-dimensional subspace, for which the prefactor in Eq.~\eqref{eq:fft_higher} vanishes. To apply the 4-dimensional Fourier transform to each data set, it suffices to send the photon once through the setup. The initial state is transformed into $\ket{\psi'} = \ffto{4}(\ket{\psi})$, whose form reads
\begin{eqnarray}
    \ket{\psi'} & = & \ffto{4}(\alpha_0 \ket{0} + \alpha_1 \ket{1} + \alpha_2 \ket{2} + \alpha_3 \ket{3}) \\
    & & + \ \ffto{4}(\beta_0 \ket{8} + \beta_1 \ket{9} + \beta_2 \ket{10} + \beta_3 \ket{11}) \\
    & = & \alpha'_0 \ket{0} + \alpha'_1 \ket{1} + \alpha'_2 \ket{2} + \alpha'_3 \ket{3} \\
    & & + \ \beta'_0 \ket{8} + \beta'_1 \ket{9} + \beta'_2 \ket{10} + \beta'_3 \ket{11},
\end{eqnarray}
where $\alpha'_j$ and $\beta'_j$ are Fourier images of data points $\alpha_j$ and $\beta_j$, respectively.

The representation in Eq.~\eqref{eq:amplenc} of a data set as a superposition of basis states corresponds to the so-called amplitude encoding \cite{schuld_supervised_2018}. A collection of $N$ data sets is then represented as a superposition of $N$ single-data-set superpositions. The processing of this larger superposition requires in our design no additional overhead in resources. This feature is akin to the proposal in Ref.~\cite{asaka_quantum_2020}, where one data set is represented as a single high-dimensional basis state using basis encoding \cite{schuld_supervised_2018} and many data sets are encoded into a superposition of such basis states. The processing of many data sets in that proposal also does not incur any additional computational costs.
}

\section{Polarization-enhanced scheme}
\label{sec:polenhanced}

\begin{figure*}
    \centering
    \includegraphics[width=\linewidth]{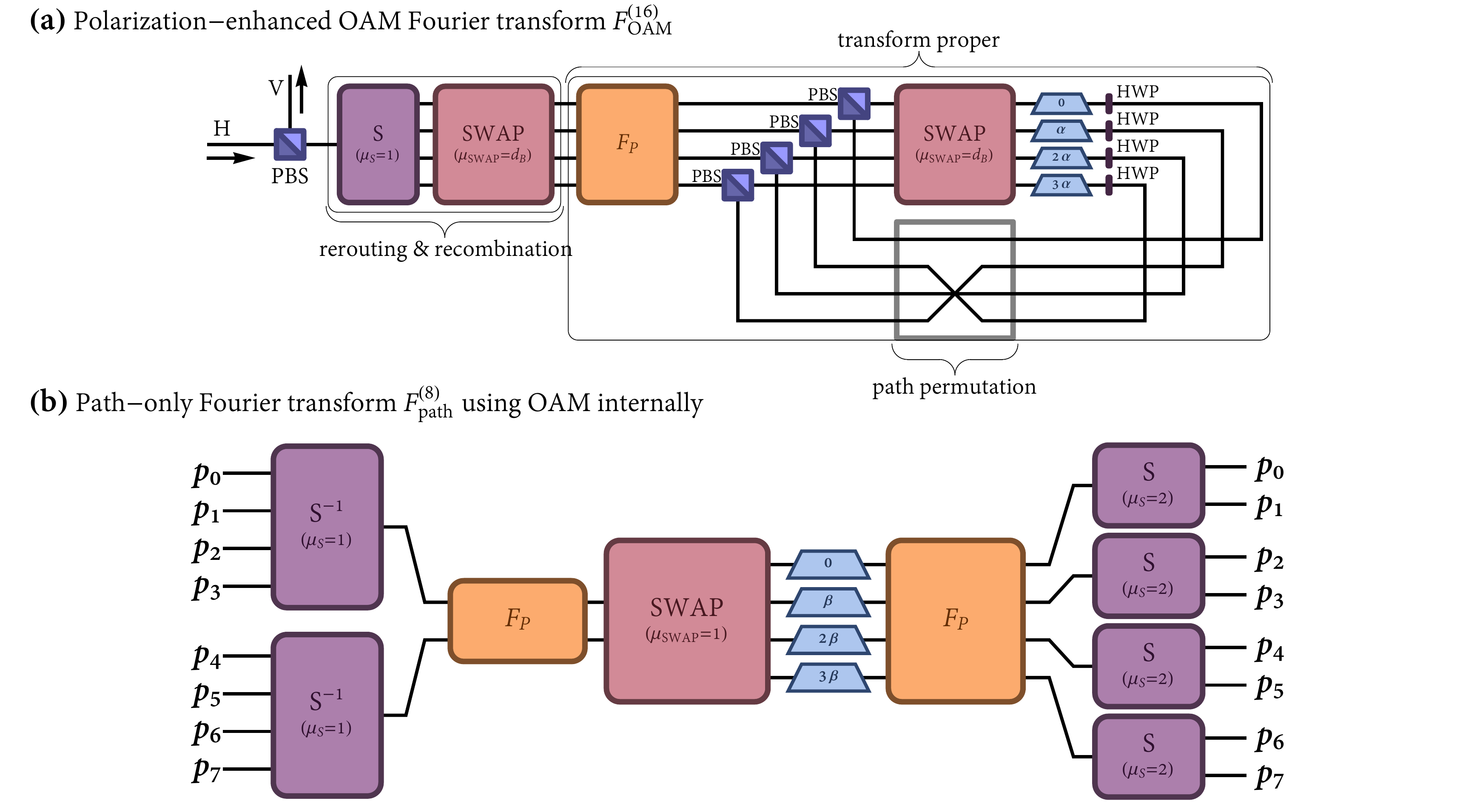}
    \caption{Modifications of the basic setup for the Fourier transform in the OAM. (a) For dimensions $d = 2^{M}$ with even $M$ one can reduce the number of elements by employing the polarization of light. Such a polarization-enhanced setup for $d = 2^4 = 16$ is shown in the figure, where the polarization of the incoming light is $H$. The propagation of all OAM eigenstates proceeds as in the basic setup until they reach half-wave plates (HWP) set to $45^\circ$ that rotate the polarization into $V$. After that, the eigenstates are rerouted back to the initial part of the setup, this time propagating backward. The resulting eigenstates leave the setup via the upper port of the first polarizing beam splitter (PBS). As the backward propagation through block $\fp$ corresponds to the inverse Fourier transform, an additional path permutation is added after half-wave plates to compensate for this effect. For details see the main text. (b) OAM-enhanced path-only Fourier transform. The first and third modules are removed from the basic setup for the OAM Fourier transform and are replaced by two series of OAM sorters. This way we obtain a $d$-dimensional path-only Fourier transform, which uses the OAM as an internal degree of freedom. In the figure a specific case for $d = 8$ is shown.}
    \label{fig:polenhanced}
\end{figure*}

As is evident from Fig.~\ref{fig:fft_paral_even_odd}, the setup for the Fourier transform displays a high degree of symmetry for dimensions $d = 2^{M}$, when the exponent $M$ is an even number. For such dimensions, the last recombination module is just an inverse of the first rerouting module and the two path-only Fourier transforms in the center module are identical. It turns out that one can remove the last module altogether as well as the second path-only Fourier transform. This can be achieved using a polarization of light as an additional degree of freedom that controls whether the light traverses the setup forward or backward. This polarization-enhanced scheme is explicitly depicted in Fig.~\ref{fig:polenhanced}(a) for dimension $d = 16$. If we fix the polarization of the incoming photon to horizontal, it passes the first part of the setup as in the original scheme. A series of half-wave plates, inserted after Dove prisms, transforms the polarization of light from horizontal to vertical. The photon then encounters a series of polarizing beam splitters that reflects it into the initial part of the setup, but now backward. The photon leaves the setup via the polarizing beam splitter prepended to the setup as shown in Fig.~\ref{fig:polenhanced}(a).

One has to bear in mind that the backward pass through the implementation $\fp$ of the path-only Fourier transform results in the inverse transformation $\fftpn^{-1}$, not $\fftpn$. This undesirable effect can be counteracted by noting that the Fourier transform $\fftpn$ in an arbitrary dimension $d_A$ satisfies the following relations \cite{PhysRevA.52.4853}
\begin{eqnarray}
    \fftpn = \fftpn^{-1} \cdot \fftpn^2,
\end{eqnarray}
where
\begin{eqnarray}
    \fftpn^2 (\ket{m}_O\ket{0}_P) & = & \ket{m}_O\ket{0}_P, \\
    \fftpn^2 (\ket{m}_O\ket{p}_P) & = & \ket{m}_O\ket{d_A - p}_P, \ 1 \leq p < d_A.
\end{eqnarray}
In other words, the square of the Fourier transform acts as a mere permutation of paths. It leaves the zeroth path unaffected and reverses the order of remaining paths. When we add an extra module to the setup, which implements this path permutation, we effectively compensate for the inversion of the path-only Fourier transform caused by the backward propagation.

As is shown in Appendix \ref{sec:other_elements}, in this polarization-enhanced scheme the number of beam splitters scales as
\begin{eqnarray}
    \numbs{F}{d} \approx \frac{7}{4}\sqrt{d}\log(d).
\end{eqnarray}
We obtain the same scaling as in the original setup, but the scaling factor is more favorable, which may prove useful in actual experimental implementations.

\section{OAM-enhanced path-only Fourier transform}
\label{sec:oamenhanced}

In cases where the path encoding is favorable, our scheme also offers advantages. It can be easily modified to act as a path-only Fourier transform, where the OAM plays the role of an intermediary that is used only inside the physical implementation \cite{[{The use of polarization as an auxiliary degree of freedom in the implementation of the path Fourier transform was reported in: }]PhysRevLett.119.080502}. The modification consists in removing the first and third modules from the setup and replacing them with a series of sorters as depicted in Fig.~\ref{fig:polenhanced}(b). This way, the new setup has $d$ input and $d$ output paths. The setting of individual modules of the setup has to reflect this modification. Unlike in the original scheme, the multiplicity of the swap operator is set to $\multswap = 1$. The settings of the subsequent components in the setup depend on the parity of $M$ in $d = 2^M$. For even $M$, the number of paths in the center module of the setup, and hence also the multiplicity of OAM eigenstates, stays constant. The angles of Dove prisms are then set such that a Dove prism in path $k$ is rotated through angle $k \beta$, where $\beta = \pi/d$, and the sorters in the third module are $d_B$-dimensional sorters with a default multiplicity $\mults=1$. For odd $M$, there is twice as many paths leaving the swap as those entering it, see a specific example for $d = 2^3 = 8$ in Fig.~\ref{fig:polenhanced}(b). As a result, the multiplicity of OAM eigenstates leaving the swap is 2. This fact is reflected in the angle of Dove prisms, where now $\beta = \pi/(2 d)$, and the multiplicity of sorters in the third module, which is set to $\mults = 2$.

The number of beam splitters in the OAM-enhanced setup scales as $4 d$ with the dimension, see Appendix \ref{sec:other_elements}. Even though the efficient scaling of the original scheme is lost, it is still an improvement over the traditional approach based on the design of Ref.~\cite{Torma1996}, which scales as $O(d \log_2(d))$. Already for $d = 2^9 = 512$ is our scheme more resource-efficient than the traditional approach. The linear scaling results from the interferometric implementation of OAM sorters \cite{Leach2002}. Nevertheless, in the present case the sorters do not have to be built using the interferometric design of Leach \textit{et al.} \cite{Leach2002} and more efficient designs can be employed, such as that of Berkhout \textit{et al.} \cite{Berkhout2010}. Using their design, we retrieve the original scaling $O(\sqrt{d}\log_2(d))$ in the number of beam splitters and other optical elements.

For dimensions $d = 2^M$ with even $M$, one can also employ polarization in the OAM-enhanced scheme in the way analogous to that in the previous section. The second part of the setup is thus removed; a series of $\sqrt{d}$ polarizing beam splitters is added to the setup and a series of $d$ polarizing beam splitters is prepended to the setup. This polarization-OAM-enhanced path-only Fourier transform requires approximately $3 d$ beam splitters. From dimension $d = 2^8 = 256$ onward, this scheme requires fewer beam splitters than the traditional approach. For details consult Appendix \ref{sec:other_elements}.

\section{Conclusion}
\label{sec:conclusion}

We present an efficient implementation of the Fourier transform acting on the orbital angular momentum of light. Our scheme works in the quantum regime with single photons as well as in the classical regime with classical light. The distinctive feature of our implementation is that the Fourier transform works in parallel on many $d$-dimensional subspaces simultaneously.

Only standard optical elements are used in the construction of the $d$-dimensional OAM Fourier transform. The number of these elements scales like $O(\sqrt{d} \log_2(d))$ as opposed to scaling $O(d)$ reported in Ref.~\cite{oamfft}. We show how a polarization of light can be utilized to reduce the number of elements even further. Moreover, using the efficient design of the OAM sorter \cite{Berkhout2010}, one can implement the Fourier transform in the path encoding with only $O(\sqrt{d} \log_2(d))$ elements as opposed to $O(d \log_2(d))$ found in Ref.~\cite{Torma1996}. In such an implementation, the OAM is an auxiliary degree of freedom, which is used only inside the setup.

\red{The favorable scaling of our design can be compared with the capabilities of the current experimental technologies. In the optical quantum domain, photonic chips represent mature and successful technology, which has been used for quantum computation tasks such as boson sampling \cite{spring_boson_2013,tillmann_experimental_2013} with path being the usual degree of freedom used in these applications. In the case of OAM on chips, considerable progress has been made 
\cite{shen_optical_2019}% \cite{shen_optical_2019,li2021optical}
, but the technology is still in its infancy \cite{chen_mapping_2018,PhysRevLett.124.153601}. At present, bulk optics represents a more promising experimental platform. The setup employed in Ref.~\cite{PhysRevLett.120.260502} used 30 interferometers, each of which manipulated single photons in three degrees of freedom including OAM. This amounts to 60 beam splitters. For comparison, our design of the 16-dimensional Fourier transform requires 47 beam splitters (or only 33 beam splitters when the polarization is utilized as in section \ref{sec:polenhanced}). Moreover, a state-of-the-art experiment has been reported recently in Ref.~\cite{Zhong1460}, where 50 polarizing beam splitters and a bulk interferometer representing 300 beam splitters were used. Such a number is more than sufficient for the construction of the 128-dimensional Fourier transform (or the 256-dimensional transform in the polarization-enhanced scheme).}

% 259 BS for d=128 and 367 BS for d=256

The Fourier transform finds many applications in the domain of classical computation. Our scheme could be used to process the classical information stored in the OAM of a light beam. It could also be utilized in photonic computational architectures, where Fourier transforms represent an essential computational primitive \cite{lpezpastor2019arbitrary, Saygin2020, pereira2020universal}. In the field of quantum tomography, mutually unbiased bases play a crucial role. One such basis is generated, when the Fourier transform is applied to the computational basis. We present an efficient implementation for dimensions of the form $d = 2^M$, for which $d + 1$ different mutually unbiased bases exist \cite{Wootters1989,Ivonovic_1981}. An interesting open question is how to adjust our setup for the generation of other mutually unbiased bases. \red{The Fourier transform possesses many remarkable algebraic properties \cite{1163843}, one of which was used in this paper. It is another open question, whether other properties can be utilized as well to reduce the complexity of the resulting implementation scheme ever further.}

\section{Acknowledgement}

The author thanks Mirjam Weilenmann for valuable comments on the manuscript. The support of Austrian Academy of Sciences and the University of Vienna via the QUESS project (Quantum Experiments on Space Scale) is acknowledged.

\bibliography{ref.bib}

\appendix

\section{OAM sorter and swap}
\label{sec:sorter_swap}

The OAM sorter is used in the setup of the Fourier transform, first, to reroute the incoming eigenstates into multiple paths and, second, to recombine all the resulting eigenstates into a single output path. The eigenstates $\ket{k}_O$ that enter the $d_A$-dimensional sorter $\sorter{d_A}$ via path $\ket{0}_P$ are transformed according to \cite{oamfft}
\begin{equation}
\sorter{d_A}(\ket{k}_O\ket{0}_P) = \ket{d_A \cdot \left\lfloor\frac{k}{d_A} \right\rfloor}_O\ket{\rule{0ex}{2.5ex}k \ \mathrm{mod} \ d_A}_P,
\label{eq:sorter_gen}
\end{equation}
where $\lfloor x \rfloor$ denotes the integer part of number $x \in \mathbb{R}$. This ``modulo property'' of the OAM sorter \cite{oamfft,Leach2002} allows for optimal redistribution of the incoming OAM eigenstates into multiple propagation modes. As a result, many OAM eigenstates propagate along the same path, which ultimately leads to efficient scaling $O(\sqrt{d}\log d)$ of the whole scheme. The number $k$ of OAM quanta in the formula \eqref{eq:sorter_gen} is unbounded. When we consider only the eigenstates $\ket{k}_O$ with $0 \leq k < d_A$, the formula reduces to
\begin{equation}
\sorter{d_A}(\ket{k}_O\ket{0}_P) = \ket{0}_O\ket{k}_P.
\label{eq:sorter_simple}
\end{equation}

The OAM sorter can be generalized into the swap operator, whose action on eigenstates $\ket{k}_O$ with $0 \leq k < d_B$ for the special case of $d_A = d_B$ reads
\begin{eqnarray}
    \swapgen{d_A, d_A}{1}\left(\ket{k}_O\ket{p}_P \right) = \ket{p}_O\ket{k}_P,
\end{eqnarray}
where $0 \leq p < d_A$ are propagation paths. In the formula above, the first, second, and third subscript stands for the number of input ports, the number of output ports, and the multiplicity of $\swap$, respectively. In this section we discuss the algebraic properties of the physical implementation of the swap operator. For its structure refer to section \ref{sec:explicit_setups}. In the setup for the Fourier transform the swaps have to act on eigenstates of the form $\ket{\mu \cdot k}$ for some fixed $\mu \in \mathbb{N}$. We call this number the \emph{multiplicity} of the eigenstates. The setup of the swap operator can be adjusted also for the case with $d_A \neq d_B$ as well as when $\mu \neq 1$. In our discussion we are interested only in a specific class of OAM eigenstates, whose multiplicity is of the form $\mu \, d_A / d_B$, where $\mu \in \mathbb{N}$. Note that we assume $d_A \geq d_B$ and take into account only dimensions that are powers of two and so $d_A/d_B$ is an integer. If we assume that the number of input paths is equal to $d_A$ and the number of output paths is $d_B < d_A$, the action of $\swap$ on eigenstates with arbitrarily large $k$ is represented by equality
\begin{multline}
    \swapgen{d_A, d_B}{\mu}\left( \ket{\mu \cdot \frac{d_A}{d_B} \cdot k}_O\ket{p}_P \right) = \\
    \ket{\mu \cdot \left( d_A \cdot \left\lfloor\frac{k}{d_B} \right\rfloor + p \right)}_O\ket{\rule{0ex}{3.2ex}k \ \mathrm{mod} \ d_B}_P.
    \label{eq:swap_gen}
\end{multline}
This action is nothing but a mere permutation of (a subset of) basis states of the composite OAM-path Hilbert space. By inspection of formula \eqref{eq:swap_gen} one sees that the multiplicity of eigenstates that leave the setup is equal to $\mu$. When we need an implementation of the swap operator with the number of input ports smaller than the number of output ports, we operate $\swap$ backward, obtaining $\swap^{-1}$. The inverse of formula \eqref{eq:swap_gen} can be rewritten into
\begin{multline}
    \swapgen{d_A, d_B}{\mu}^{-1}\left( \ket{\mu \cdot k}_O\ket{p}_P \right) = \\
    \ket{\mu \cdot \frac{d_A}{d_B} \cdot \left( d_B \cdot \left\lfloor\frac{k}{d_A} \right\rfloor + p \right)}_O\ket{\rule{0ex}{3.2ex}k \ \mathrm{mod} \ d_A}_P.
    \label{eq:swap_gen_inv}
\end{multline}
When we set the multiplicity to $\mu = d_B$ and consider only eigenstates $\ket{k}_O$ with $0 \leq k < d_A$, formula \eqref{eq:swap_gen_inv} above reduces to
\begin{multline}
    \swapgen{d_A, d_B}{d_B}^{-1}\left(\ket{d_B \cdot k}_O\ket{p}_P \right) = \ket{d_A \cdot p}_O\ket{k}_P.
    \label{eq:swap_usef}
\end{multline}
This relation is used in the second step \eqref{eq:derivation_swap} of the decomposition presented in section \ref{sec:derivation}.

In the setup of the OAM Fourier transform, a sorter forms a module with a swap operator. The first module in the setup affects the incoming OAM eigenstates in the following way:
\begin{eqnarray}
     & \text{input:} & \ket{q}_O \ket{0}_P \\
     & \xrightarrow{\sorter{d_A}} & \ket{d_A \cdot \left\lfloor\frac{q}{d_A} \right\rfloor}_O\ket{\rule{0ex}{2.5ex}q \ \mathrm{mod} \ d_A}_P \\
     & \xrightarrow{\swap{d_A, d_B}{d_B}} & \ket{d_B \cdot (q \ \mathrm{mod} \ d_A)}_O\ket{\left\lfloor\frac{q}{d_A} \right\rfloor}_P. \label{eq:sort_and_swap_orig}
\end{eqnarray}
This formula has a very simple interpretation, which is depicted in Fig.~\ref{fig:evolution}(b).

\section{Decomposition of the Fourier transform}
\label{sec:derivation}

The entire setup for the OAM Fourier transform consists of three modules. The first module, comprising a sorter and a swap operator, reroutes different OAM eigenstates to different paths. This module is discussed in the previous section. The second module is the Fourier transform proper and the third module recombines all eigenstates into a single output path. Its structure mirrors that of the first module. Let us study first the evolution of the mode $\ket{d_B \, m}_O \ket{l}_P$ with some $m, l \in \mathbb{N}$, when it propagates through the second module of the setup, cf. Eq.~\eqref{eq:main_eq} and Fig.~\ref{fig:evolution}. Unlike in Eq.~\eqref{eq:main_eq} in the derivation below we already take into account also the multiplicities of OAM eigenstates when they propagate through the physical setup, see Fig.~\ref{fig:fft_paral_even_odd}. The incoming mode undergoes individual steps of its evolution as follows:
\begin{eqnarray}
     & \text{input:} & \ket{d_B \, m}_O \ket{l}_P \\
     & \xrightarrow{\fp} & \ket{d_B \, m}_O \frac{1}{\sqrt{d_B}} \sum_{j=0}^{d_B-1} e^{\frac{2 \pi i}{d_B} j l} \ket{j}_P \\
     & \xrightarrow{\swap^{-1}} & \frac{1}{\sqrt{d_B}} \sum_{j=0}^{d_B-1} e^{\frac{2 \pi i}{d_B} j l} \ket{d_A \, j}_O \ket{m}_P \label{eq:derivation_swap} \\
     & \xrightarrow{\mathrm{Dove}} & \frac{1}{\sqrt{d_B}} \sum_{j=0}^{d_B-1} e^{\frac{2 \pi i}{d_B} j l + \frac{2 \pi i}{d} j m} \ket{d_A \, j}_O \ket{m}_P \\
     & \xrightarrow{\fp} & \frac{1}{\sqrt{d}} \sum_{j=0}^{d_B-1}\sum_{k=0}^{d_A-1} e^{\frac{2 \pi i}{d_B} j l + \frac{2 \pi i}{d} j m + \frac{2 \pi i}{d_A} k m} \ket{d_A \, j}_O \ket{k}_P. \nonumber
\end{eqnarray}
It is easy to check that the exponential in the last expression can be rewritten as
\begin{equation}
    e^{\frac{2 \pi i}{d_B} j l + \frac{2 \pi i}{d} j m + \frac{2 \pi i}{d_A} k m} = e^{\frac{2 \pi i}{d} (d_A l + m)(d_B k + j)}.
\end{equation}
We thus obtain the transformation rule
\begin{multline}
    \ket{d_B \, m}_O \ket{l}_P \to \\ \frac{1}{\sqrt{d}} \sum_{j=0}^{d_B-1}\sum_{k=0}^{d_A-1} e^{\frac{2 \pi i}{d} (d_A l + m)(d_B k + j)} \ket{d_A \, j}_O \ket{k}_P. \label{eq:fft_theory}
\end{multline}
The purpose of the first module of the setup is then just to transform the incoming eigenstate $\ket{q}_O \ket{0}_P$ into the form $\ket{d_B \, m}_O \ket{l}_P$ and analogously the role of the third module of the setup is to transform individual terms $\ket{d_A \, j}_O \ket{k}_P$ in the sum above into the form $\ket{r}_O \ket{0}_P$.

As follows from \eqref{eq:sort_and_swap_orig}, the first module of the setup implements the transformation
\begin{equation}
    \ket{q}_O \ket{0}_P \to \ket{d_B \, (q \bmod d_A)}_O \ket{\left\lfloor \frac{q}{d_A} \right\rfloor}_P. \label{eq:sort_and_swap}
\end{equation}
We can identify the indices in Eqs.~\eqref{eq:fft_theory} and \eqref{eq:sort_and_swap} as $m \coloneqq (q \bmod d_A)$ and $l \coloneqq \left\lfloor q/d_A \right\rfloor$. As a result, $q = d_A \, l + m$. The action of the first and second modules of the setup then reads
\begin{equation}
    \ket{q}_O \ket{0}_P \to \frac{1}{\sqrt{d}} \sum_{j=0}^{d_B-1}\sum_{k=0}^{d_A-1} e^{\frac{2 \pi i}{d} q (d_B k + j)} \ket{d_A \, j}_O \ket{k}_P.
\end{equation}

Analogously, the third module of the setup performs the operation $\ket{d_A \, j}_O \ket{k}_P \to \ket{r}_O \ket{0}_P$, where $r \coloneqq d_B \, k + j$. The whole setup thus acts like
\begin{equation}
    \ket{q}_O \ket{0}_P \to \frac{1}{\sqrt{d}} \sum_{r=0}^{d-1} e^{\frac{2 \pi i}{d} q r} \ket{r}_O \ket{0}_P,
\end{equation}
which corresponds exactly to the $d$-dimensional Fourier transform of the OAM of the incoming photon.

\section{Optimal choice of $d_A$ and $d_B$}
\label{sec:optimal_choice}

The values of $d_A$ and $d_B$ in the decomposition formula \eqref{eq:main_eq} are not a priori fixed. They only have to satisfy $d = d_A \, d_B$. In this section we derive the optimal values for $d_A$ and $d_B$ when we take the total number of beam splitters in the setup as our figure of merit. This choice of the cost function is motivated by the fact that the number of beam splitters correlates with the interferometric complexity of the setup.

The number of beam splitters necessary for the implementation of the $d$-dimensional sorter equals $\numbs{\sorter{}}{d} = 2 (d - 1)$ \cite{oamfft}. Similarly, for the path-only Fourier transform we obtain $\numbs{pF}{d} = (d \log_2 d)/2$ \cite{Torma1996}. To implement the swap operator for the input dimension $\din$ and the output dimension $\dout$ one first decides which of the two dimensions is larger and sets $\dmin = \min (\din, \dout)$ and $\dmax = \max (\din, \dout)$. The number of beam splitters in the implementation of the swap operator is then given by \cite{oamfft}
\begin{multline}
    \numbsA{\swap}{\din}{\dout} = \frac{1}{2} \dmax \log_2 \dmax + \\ \dmin \log_2 \dmin + \dmax - 2 \dmin + 1. \label{eq:num_swap}
\end{multline}
The total number of beam splitters required to implement the OAM Fourier transform in dimension $d = d_A \, d_B$ is then
\begin{eqnarray}
    \numbsA{F}{d_A}{d_B} & = & \numbs{\sorter{}}{d_A} + \numbs{pF}{d_A} \nonumber \\
    & + & \numbs{\sorter{}}{d_B} + \numbs{pF}{d_B} \label{eq:total_nbs} \\
    & + & 3 \, \numbsA{\swap}{d_A}{d_B}. \nonumber
\end{eqnarray}
The expression above is symmetric with respect to the exchange of $d_A$ and $d_B$. Without loss of generality we can thus investigate only the cases with $d_A \geq d_B$ as was done in the main text. Since $d_A = d/d_B$, we study the values of $d_B$ for which $1 \leq d_B \leq \sqrt{d}$. Moreover, $\dmin = d_B$ and $\dmax = d_A$ in Eq.~\eqref{eq:num_swap}. As we treat only dimensions that are powers of two, we can set $\dmin = 2^m$ and $\dmax = 2^{M-m}$, where $d = 2^M$ and $1 \leq m \leq \lfloor M/2 \rfloor$. From \eqref{eq:total_nbs} one obtains the total number of beam splitters
\begin{multline}
    \numbsA{F}{2^{M-m}}{2^m} = 2^m \, \left(\frac{7}{2} \, m - 4 \right) - 1 \\ + 2^{M-m} \, (2(M-m) + 5).
\end{multline}

In the following we show that this number is minimal, when $m$ attains its maximum possible value, i.e. $m = \lfloor M/2 \rfloor$. Let us define $g(m) \coloneqq \numbsA{F}{2^{M-m}}{2^m}$. It is easy to check that
\begin{multline}
    g(m) - g(m-1) = 2^{m-2} \left( 7 \, m - 1 \right) \\ - 2^{M-m} \left(2(M-m) + 9 \right).
\end{multline}
This expression can be bounded from above like
\begin{multline}
    g(m) - g(m-1) \leq 2^{\left\lfloor \frac{M}{2} \right\rfloor - 2} \left( 7 \, \left\lfloor \frac{M}{2} \right\rfloor - 1 \right) \\
    - \ 2^{M-\left\lfloor \frac{M}{2} \right\rfloor} \left(2 \left( M-\left\lfloor \frac{M}{2} \right\rfloor \right) + 9 \right).
    \label{eq:gdiff}
\end{multline}
When $M$ is even, i.e. $M = 2 \, K$ for some integer $K$, the expression on the right-hand side reduces to $- 2^{K-2}(K + 37)$, which is evidently negative. The function $g$ is therefore monotonically decreasing and in turn assumes its minimum for the largest allowed argument $m = \lfloor M/2 \rfloor$. For odd $M$, i.e. $M = 2K+1$, the right-hand side of \eqref{eq:gdiff} reduces to $-2^{K-2}(9K+89)$, which is also negative, and the function $g$ thus attains its minimum also for $m = \lfloor M/2 \rfloor$.

To conclude, the optimal choice of $d_A$ and $d_B$ for a fixed dimension $d = 2^M$ reads
\begin{equation}
    d_A = 2^{M - \left\lfloor \frac{M}{2} \right\rfloor}, \quad d_B = 2^{\left\lfloor \frac{M}{2} \right\rfloor}. \label{eq:optimum}
\end{equation}
It turns out that we would come to the same conclusion if we chose not the number of beam splitters, but instead the number of phase shifters or the number of holograms as our figure of merit. As for the number of Dove prisms, the choice \eqref{eq:optimum} is optimal unless the dimension is too large. Specifically, it is optimal for $K \leq 6$ when $d = 2^{2 K}$, i.e. $d \leq 2^{12} = 4096$; and for $K \leq 12$ when $d = 2^{2 K + 1}$, i.e. $d = 2^{25} \approx 33.5 \times 10^6$.

\section{Numbers of optical elements}
\label{sec:other_elements}

\begin{figure}
    \centering
    \includegraphics[width=0.9\linewidth]{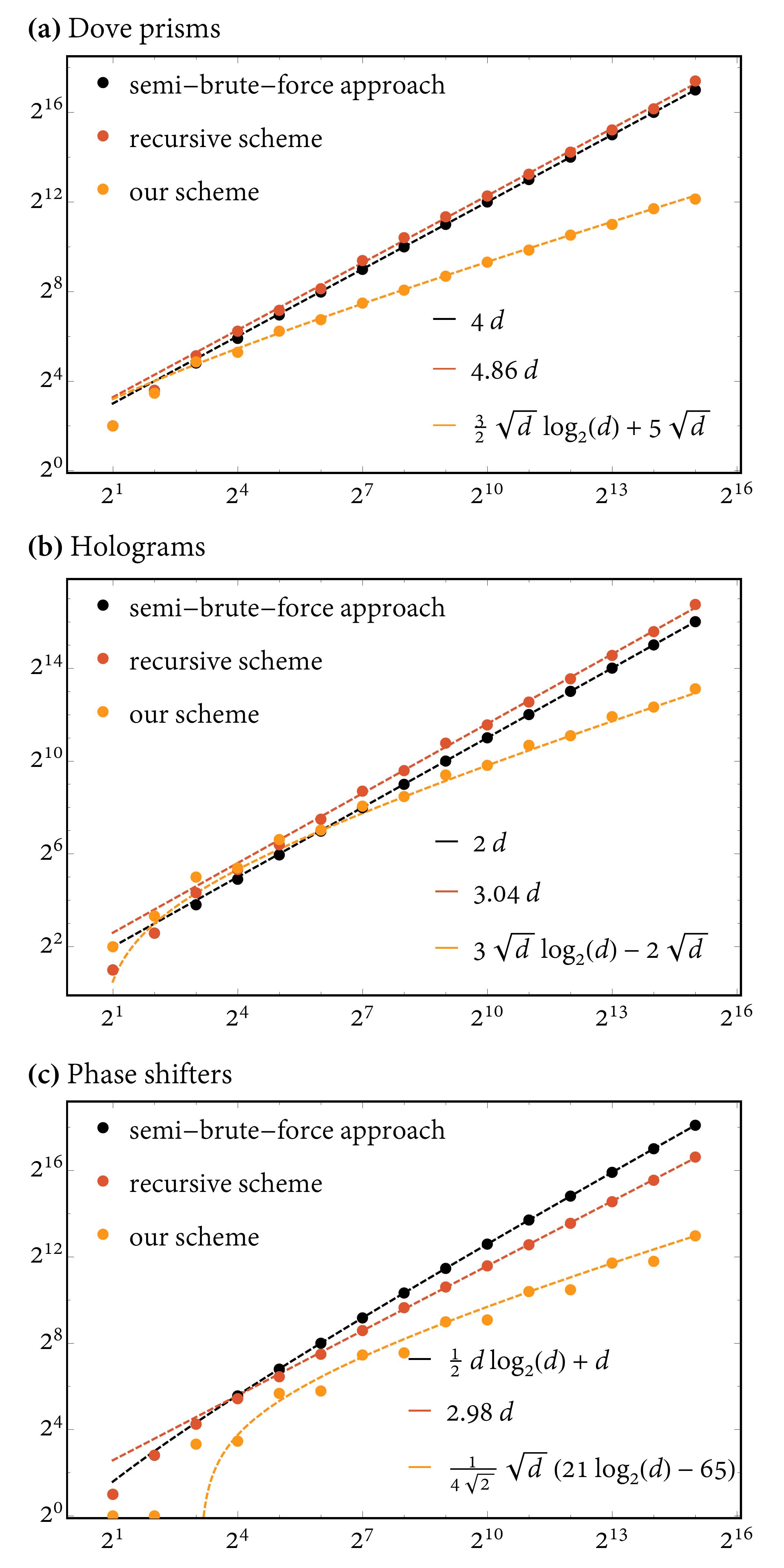}
    \caption{The numbers of (a) Dove prisms, (b) holograms, and (c) phase shifters in the setup for the OAM Fourier transform as functions of the dimension $d$, using three different approaches. Both, the semi-brute-force and recursive approaches require asymptotically considerably more resources than our scheme. Note that both axes in the plots use logarithmic scaling. The point markers denote precise numbers, whereas lines represent the growing trends analytically. For more details see the text.}
    \label{fig:comparison_elems}
\end{figure}

In the previous section we derived the optimal choices for $d_A$ and $d_B$. In this section we present the corresponding numbers of optical elements that compose the setup. For even $M$ the formula \eqref{eq:optimum} reduces to $d_A = d_B = 2^{M/2}$. In such a case the total number of beam splitters reads
\begin{equation}
    \numbsA{F}{2^{\frac{M}{2}}}{2^{\frac{M}{2}}} = 2^{\frac{M}{2}} \, \left(\frac{11}{4} \, M + 1 \right) - 1.
\end{equation}
This number scales for large $M$ as $(11/4)\sqrt{d}\log_2(d) = 2.75 \sqrt{d}\log_2(d)$. For odd $M$ the optimal scenario corresponds to $d_A = 2^{(M+1)/2}$ and $d_B = 2^{(M-1)/2}$, for which the total number of beam splitters is equal to
\begin{multline}
    \numbsA{F}{2^{\frac{M+1}{2}}}{2^{\frac{M-1}{2}}} = 2^{\frac{M-1}{2}} \, \left(\frac{15}{4} \, M + \frac{25}{4} \right) - 1.
\end{multline}
This expression scales as $(15/4\sqrt{2})\sqrt{d}\log_2(d) \approx 2.65 \sqrt{d}\log_2(d)$. For both even and odd $M$ the optimal number of beam splitters in our implementation thus scales as
\begin{equation}
    \numbs{F}{d} \sim \frac{11}{4} \sqrt{d} \, \log_2(d).
\end{equation}
This relation is one of the main results of this work. Let us study now the number of other optical elements present in the setup of the Fourier transform.

Very similar discussion can be done also for Dove prisms, holograms, phase shifters, and mirrors. (Refer to section \ref{sec:explicit_setups} for the explicit structure of individual building blocks of our scheme.) One obtains the following scaling properties
\begin{eqnarray*}
    \numdove{F}{d} & = & 3 d_B \log_2 d_B + 9 d_A - 4 d_B - 5, \\
    \numholo{F}{d} & = & 3 d_A \log_2 d_A + 3 d_B \log_2 d_B + 2 d_A - 4 d_B + 2, \\
    \numphas{F}{d} & = & 5 d_A \log_2 d_A + \frac{1}{2} d_B \log_2 d_B - 10 d_A - d_B + 11, \\
    \nummirr{F}{d} & = & 7 d_A \log_2 d_A + 4 d_B \log_2 d_B - 8 d_A + 11 d_B - 3.
\end{eqnarray*}

For dimensions with even $M$, i.e. $d = 2^{2 K}$, these formulas attain the form
\begin{eqnarray}
    \numdove{F}{d} & = & \sqrt{d} \ \left(\frac{3}{2} \log_2(d) + 5 \right) - 5 \nonumber \\
    & = & \sqrt{d} \ (1.5 \log_2(d) + 5) - 5, \\
    \numholo{F}{d} & = & \sqrt{d} \ (3 \log_2(d) - 2) + 2, \\
    \numphas{F}{d} & = & \sqrt{d} \ \left(\frac{11}{4} \log_2(d) - 11 \right) + 11 \nonumber \\
    & = & \sqrt{d} \ (2.75 \log_2(d) - 11) + 11, \\
    \nummirr{F}{d} & = & \sqrt{d} \ \left(\frac{11}{2} \log_2(d) + 3 \right) - 3 \nonumber \\
    & = & \sqrt{d} \ (5.5 \log_2(d) + 3) - 3.
\end{eqnarray}
Similarly, for dimensions with odd $M$, i.e. $d = 2^{2K+1}$, one gets
\begin{eqnarray}
    \numdove{F}{d} & = & \sqrt{d} \ \left(\frac{3}{2\sqrt{2}} \log_2(d) + \frac{25}{2\sqrt{2}} \right) - 5 \nonumber \\
    & \approx & \sqrt{d} \ (1.06 \log_2(d) + 8.84) - 5, \\
    \numholo{F}{d} & = & \sqrt{d} \ \left(\frac{9}{2\sqrt{2}} \log_2(d) + \frac{3}{2\sqrt{2}} \right) + 2 \nonumber \\
    & \approx & \sqrt{d} \ (3.18 \log_2(d) + 1.06) + 2, \\
    \numphas{F}{d} & = & \sqrt{d} \ \left(\frac{21}{4\sqrt{2}} \log_2(d) - \frac{65}{4\sqrt{2}} \right) + 11 \nonumber \\
    & \approx & \sqrt{d} \ (3.71 \log_2(d) - 11.49) + 11, \\
    \nummirr{F}{d} & = & \sqrt{d} \ \left(\frac{15}{2\sqrt{2}} \log_2(d) + \frac{37}{2\sqrt{2}} \right) - 3 \nonumber \\
    & \approx & \sqrt{d} \ (5.30 \log_2(d) + 13.08) - 3.
\end{eqnarray}
By inspection of all the formulas above we can conclude that not only beam splitters, but actually all the other relevant optical elements in the setup scale efficiently according to $O(\sqrt{d}\log_2(d))$. In Fig.~\ref{fig:comparison_elems} the exact numbers of these optical elements are plotted and compared to the numbers obtained using two alternative approaches. The semi-brute-force approach is to transform the OAM of light into the path encoding and apply a path-only Fourier transform, which is built utilizing an efficient design of Ref.~\cite{Torma1996}. The recursive approach was introduced in Ref.~\cite{oamfft} and scales better than the semi-brute-force approach. As is evident from the plots, the current scheme offers considerable improvements over these two approaches. The number of mirrors discussed above takes into account mirrors that are used to construct individual building blocks of the scheme, see section \ref{sec:explicit_setups}. This number is only a rough estimate as in the real implementation additional mirrors are usually necessary.  

Note that additional minor savings in resources can be made in our scheme. For instance, in the calculations above we took two holograms per each OAM exchanger appearing in OAM sorters. In a real setup, only one hologram is actually necessary as only the upper input path of the exchangers is used \cite{oamfft}.

The polarization-enhanced scheme, presented in section \ref{sec:polenhanced} in the main text, allows one to further reduce the number of optical elements. This improvement is possible only for dimensions $d = 2^{M}$ with even $M$, for which we fix $d_A = d_B = 2^{\frac{M}{2}} = \sqrt{d}$. For beam splitters and polarizing beam splitters together we get
\begin{eqnarray}
    \numbs{pol-F}{d} & = & \numbs{\sorter{}}{\sqrt{d}} + \numbs{pF}{\sqrt{d}} + 1\nonumber \\
    & + & 2 \, \numbsA{\swap}{\sqrt{d}}{\sqrt{d}} + \sqrt{d} \nonumber \\
    & = & \frac{7}{4}\sqrt{d} \log_2(d) + \sqrt{d} + 1. \label{eq:total_nbs_pol}
\end{eqnarray}
The scaling $O(\sqrt{d}\log_2(d))$ stays the same as in the original setup, but the improved scaling factor $7/4 = 1.75$ could be of importance in a real experimental implementation.

Very similar discussion of the number of elements can be done also for the OAM-enhanced path-only Fourier transform. The general formula changes into
\begin{eqnarray}
    \numbsA{e-F}{d_A}{d_B} & = & d_B \numbs{\sorter{}}{d_A} + \numbs{pF}{d_A} \nonumber \\
    & + & d_A \numbs{\sorter{}}{d_B} + \numbs{pF}{d_B} \label{eq:total_nbs_path} \\
    & + & \numbsA{\swap}{d_A}{d_B}. \nonumber
\end{eqnarray}
The number of beam splitters is also in this scenario minimal when the values of $d_A$ and $d_B$ satisfy relations \eqref{eq:optimum}. The resulting formulas for $d = 2^{2K}$ and $d = 2^{2K+1}$ read respectively
\begin{eqnarray}
    \numbs{e-F}{d} & = & 4d+\sqrt{d} \ \left(\frac{5}{4}\log_2(d)-5 \right)+1, \nonumber \\
    \numbs{e-F}{d} & = & 4d+\sqrt{d} \ \left(\frac{7}{4\sqrt{2}}\log_2(d)-\frac{23}{4\sqrt{2}} \right)+1. \nonumber
\end{eqnarray}
Both of these expressions scale like $O(d)$. Even though the appealing scaling of the original OAM setup is lost, it is still a better result than that reported in Ref.~\cite{oamfft}. Our OAM-enhanced scheme requires fewer beam splitters than the traditional design \cite{Torma1996} when the dimension is equal to or larger than $d = 2^9 = 512$, as opposed to $d = 2^{13} = 8192$ for the equivalent setup in Ref.~\cite{oamfft}. The linear term in formulas above is due to the interferometric implementation of the OAM sorters. However, in the case of OAM-enhanced path-only Fourier transform, the OAM sorters can be implemented using different designs, such as the one where only two plates with special profiles are used \cite{Berkhout2010}. Then even this scheme preserves the favorable scaling $O(\sqrt{d}\log_2(d))$.

One can also consider the OAM-enhanced path-only Fourier transform for dimensions $d = 2^{2K}$, where the polarization is used in a way described in section \ref{sec:polenhanced} in the main text. Such a scheme still has a linear scaling. Specifically 
\begin{eqnarray}
    \numbs{e-F}{d} & = & 3d+\sqrt{d} \ (\log_2(d)-2)+1.
\end{eqnarray}
This scheme requires less (nonpolarizing and polarizing) beam splitters than the traditional approach for $d = 2^8 = 256$ onward.

\section{Losses}
\label{sec:losses}

\begin{figure}
    \centering
    \includegraphics[width=\linewidth]{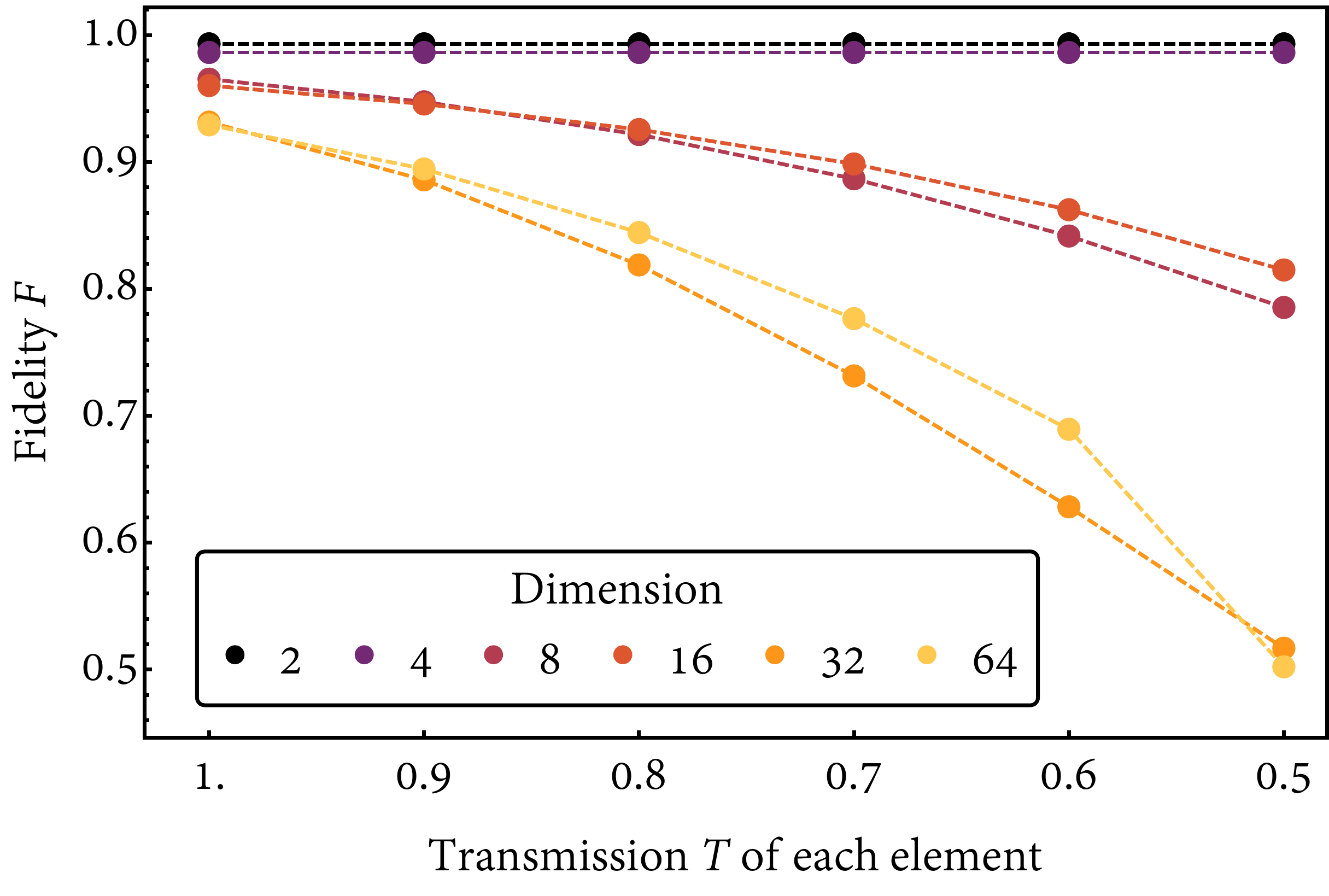}
    \caption{\red{Fidelity $F(M,U)$ between the imperfect Fourier transform $M$ and its perfect counterpart $U$ plotted as a function of the transmission $T$ of individual optical elements. For details see the text.}}
    \label{fig:losses}
\end{figure}

\red{In this section, we provide rough estimates of how much the performance of the setup of the Fourier transform is affected by imperfect operation of optical elements. There are different sources of imperfections, such as nonideal transmission of optical elements, beam distortion due to Dove prisms, varying splitting ratio of beam splitters, or suboptimal conversion efficiency of holograms. We refer to all these mechanisms as losses and quantify them by the effective transmission $T$ of each optical element. As the mirrors can be produced with very high quality and phase shifters can be implemented as mere path length differences in interferometers, we neglect losses incurred by these two types of elements. For simplicity, we also assume that all OAM modes suffer the same amount of loss and we set the effective transmission of all holograms to 90 percent. Vortex plates with such efficiencies are commercially available. For all the other elements we assume that each of them incurs the same loss, quantified by $T$.

In accordance with Ref.~\cite{ClementsScheme} we identify two kinds of losses --- the balanced losses decrease the overall efficiency without affecting the fidelity of resulting states and the unbalanced losses lead to fidelity deterioration. We focus on the latter, which result mainly from asymmetric geometry of the setup. With the aforementioned assumptions, we calculate for a given dimension $d$ the matrix $M$ corresponding to the resulting transformation and compare it with the matrix of ideal Fourier transform $U$. Following Refs.~\cite{ClementsScheme,Kumar_2021}, we quantify the effect of (unbalanced) losses on the final transformation by the normalized fidelity $F$ between $M$ and $U$, which is defined as $F = |\mathrm{Tr}(U^\dagger M)/\sqrt{d \, \mathrm{Tr}(M^\dagger M)}|^2$. The results of numerical simulations for several low dimensions are plotted in Fig.~\ref{fig:losses}. As can be seen from the plot, the fidelity curves are approximately identical for pairs of successive dimensions $d = 2^{2K-1}$ and $d = 2^{2K}$. This is caused by the fact that the structure of the setup for dimensions with even exponent is more symmetric than that for dimensions with odd exponent, cf. Fig.~\ref{fig:fft_paral_even_odd}, and the losses are thus more evenly distributed. As a result, there is no significant change in fidelity when increasing the dimension from $d = 2^{2K-1}$ to $d = 2^{2K}$. Apart from that, the losses influence the fidelity in our model in a way, which is comparable to the role of losses in rectangular decomposition of general unitaries studied in Ref.~\cite{Kumar_2021}.}

\section{Explicit setups for low dimensions}
\label{sec:explicit_setups}

\begin{figure}
    \centering
    \includegraphics[width=0.9\linewidth]{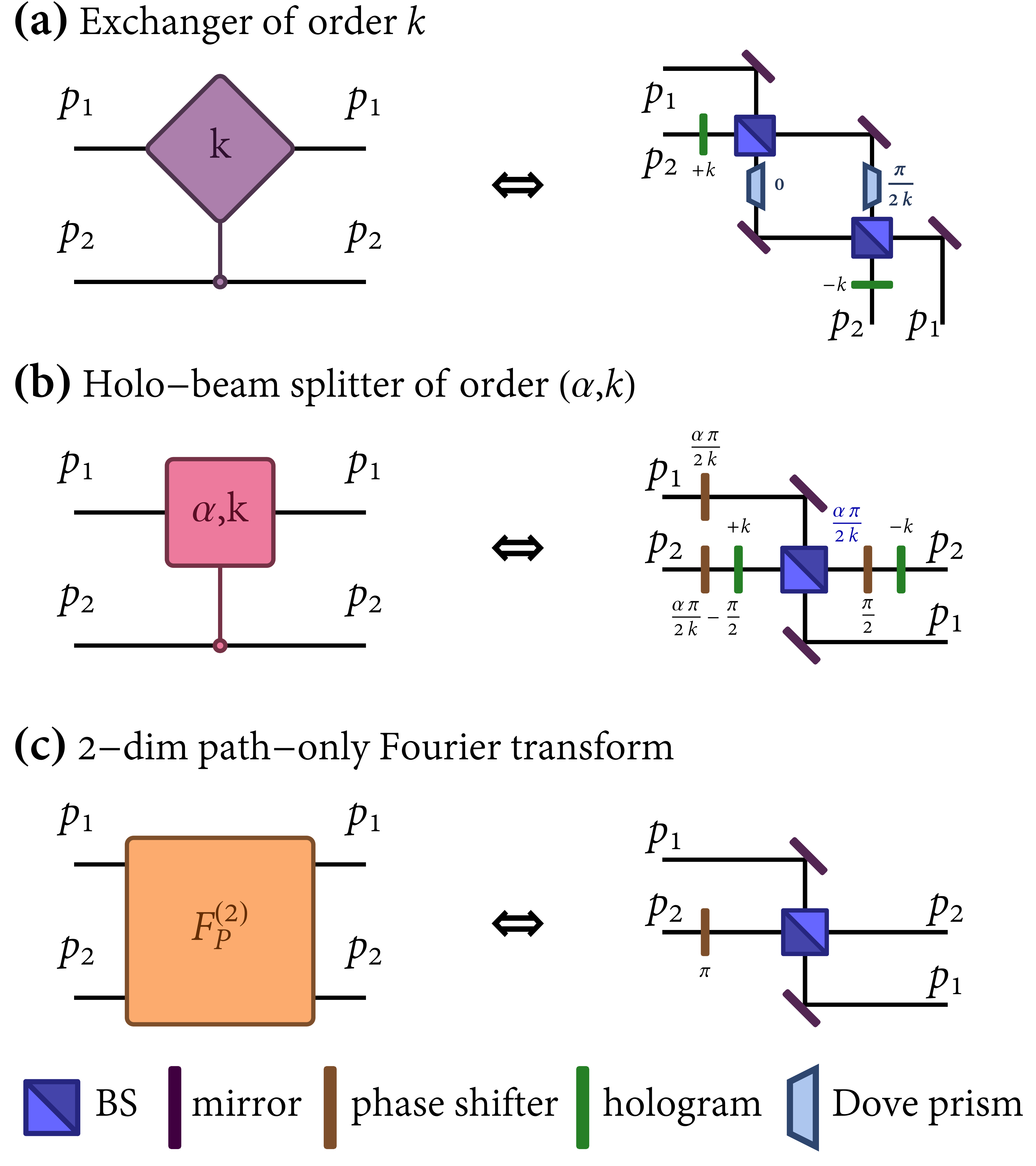}
    \caption{Building blocks of our scheme. (a) The OAM exchanger of order $k$ built as a Mach-Zehnder interferometer with two symmetric beam splitters (BS). (b) The holo-beam splitter of order $(\alpha, k)$ built using an asymmetric beam splitter with the splitting ratio equal to $\alpha \pi/(2k)$. (c) The 2-dimensional path-only Fourier transform. The undesirable side effect of a beam splitter is that it reverses the sign of OAM reflected off its interface. To correct for this inversion, additional mirrors are used.}
    \label{fig:components}
\end{figure}

\begin{figure*}
    \centering
    \includegraphics[width=\linewidth]{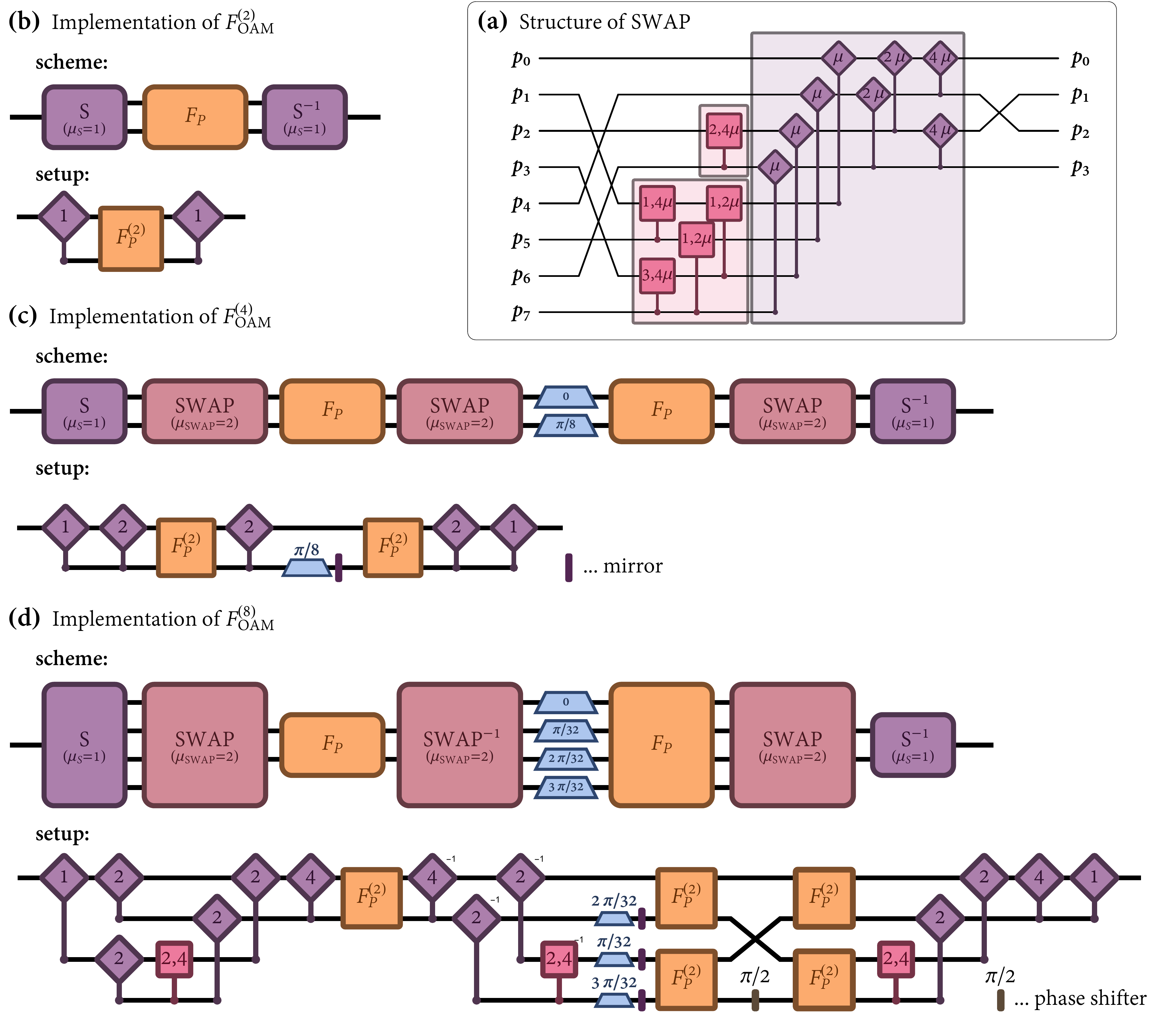}
    \caption{Explicit forms of the swap operator and setups for dimensions $d = 2$, $d = 4$, and $d = 8$. (a) The general structure of the swap operator, exemplified for the case with 8 input and 4 output ports. A swap with $\din = 2^I$ input paths and $\dout = 2^O$ output paths for some $I, O \in \mathbb{N}$ is realized as an optical network of two-input two-output optical elements. It consists of a single block of OAM exchangers and a series of blocks of holo-beam splitters. The path permutations in front of the network and after it are necessary for the correct operation of the swap, but in the resulting setup cancel out with identical permutations coming from OAM sorters and path-only Fourier transforms. The order of each exchanger and holo-beam splitter has to be multiplied by the multiplicity $\mu$ of OAM eigenstates leaving the swap. (b) The 2-dimensional OAM Fourier transform implemented using elements from Fig.~\ref{fig:components}. (c) The 4-dimensional OAM Fourier transform. An additional mirror, shown explicitly in the figure, is necessary to revert the sign-inversion of the OAM imparted by the Dove prism \eqref{eq:dove}. (d) The 8-dimensional OAM Fourier transform. The inverse signs above some elements mean that those elements are to be operated backward. Note the permuted order of Dove prisms, which is a remnant of the path permutations in the swap and the path-only Fourier transform.
    }
    \label{fig:explicit_setups}
\end{figure*}

In this final section, the explicit forms of individual building blocks of the scheme as well as several explicit setups are presented. The elementary building elements include Dove prisms, holograms, phase shifter, and mirrors. They act on the incoming OAM eigenstate $\ket{k}_O$ in the following way:
\begin{eqnarray}
    \text{Dove prism:} \quad \ket{k}_O & \to & -e^{2 i \, \alpha \, k} \ket{-k}_O, \label{eq:dove} \\
    \text{hologram:} \quad \ket{k}_O & \to & \ket{k + m}_O, \\
    \text{phase shifter:} \quad \ket{k}_O & \to & e^{i \varphi} \ket{k}_O, \\
    \text{mirror:} \quad \ket{k}_O & \to & -\ket{-k}_O,
\end{eqnarray}
where $\alpha$, $m$, and $\varphi$ are fixed and given by the form of the Dove prism, hologram, and the phase shifter, respectively. A beam splitter acts on the incoming eigenstates entering one of the two input ports $p_1$ or $p_2$ like
\begin{eqnarray}
    \ket{k}_O\ket{p_1}_P & \xrightarrow{\bs{}} & \frac{1}{\sqrt{2}}(\ket{k}_O\ket{p_1}_P + \ket{-k}_O\ket{p_2}_P), \\
    \ket{k}_O\ket{p_2}_P & \xrightarrow{\bs{}} & \frac{1}{\sqrt{2}}(\ket{-k}_O\ket{p_1}_P - \ket{k}_O\ket{p_2}_P).
\end{eqnarray}
From two beam splitters an OAM exchanger can be built \cite{oamfft}. The OAM exchanger of order $m$ is a two-input two-output interferometric device that acts on incoming eigenstates like
\begin{eqnarray*}
    \ket{k}_O\ket{p_1}_P & \xrightarrow{\exch{m}} & e^{-\frac{i \pi k}{2 m}} \cos \left(\frac{\pi k}{2 m}\right) \ket{k}_O\ket{p_1}_P \\
    & + & i \, e^{-\frac{i \pi k}{2 m}} \sin \left(\frac{\pi k}{2 m}\right) \ket{k-m}_O\ket{p_2}_P, \\
    \ket{k}_O\ket{p_2}_P & \xrightarrow{\exch{m}} & e^{-\frac{i \pi k}{2 m}} \cos \left(\frac{\pi k}{2 m}\right) \ket{k+m}_O\ket{p_1}_P \\
    & + & i \, e^{-\frac{i \pi k}{2 m}} \sin \left(\frac{\pi k}{2 m}\right) \ket{k}_O\ket{p_2}_P.
\end{eqnarray*}
The implementation of the OAM exchanger is shown in Fig.~\ref{fig:components}(a). Another important component in our scheme is a holo-beam splitter \cite{oamfft}. The holo-beam splitter of order $(\alpha, m)$ acts on incoming eigenstates like
\begin{eqnarray*}
    \ket{k}_O\ket{p_1}_P & \xrightarrow{\hbs{\alpha}{m}} & e^{\frac{i \pi \alpha }{2 m}} \cos \left(\frac{\pi \alpha }{2 m}\right) \ket{k}_O\ket{p_1}_P \\
    & - & i \, e^{\frac{i \pi \alpha }{2 m}} \sin \left(\frac{\pi \alpha}{2 m}\right) \ket{k-m}_O\ket{p_2}_P, \\
    \ket{k}_O\ket{p_2}_P & \xrightarrow{\hbs{\alpha}{m}} & e^{\frac{i \pi \alpha }{2 m}} \cos \left(\frac{\pi \alpha }{2 m}\right) \ket{k}_O\ket{p_2}_P \\
    & - & i \, e^{\frac{i \pi \alpha }{2 m}} \sin \left(\frac{\pi \alpha}{2 m}\right) \ket{k+m}_O\ket{p_1}_P.
\end{eqnarray*}
The implementation of the holo-beam splitter is shown in Fig.~\ref{fig:components}(b). Note that the action of exchangers and holo-beam splitters is slightly different from that of Ref.~\cite{oamfft}. The last elementary building block is a 2-dimensional path-only Fourier transform, whose implementation is depicted in Fig.~\ref{fig:components}(c).

More complex structures can be built from the aforementioned blocks. Specifically, OAM sorters, swap operators, and high-dimensional path-only Fourier transforms. The structure of an OAM sorter and a swap operator is demonstrated in Fig.~3 and Fig.~4 in Ref.~\cite{oamfft}, respectively.  Thanks to an optimized form of OAM exchangers and holo-beam splitters in Fig.~\ref{fig:components}, no additional Dove prisms are necessary in the construction of the swap operator in this paper (cf. Fig.~7 in Ref.~\cite{oamfft}). For convenience, its structure for a specific case of 8 input ports and 4 output ports is shown in Fig.~\ref{fig:explicit_setups}(a). When the number of input ports $\din$ and the number of output ports $\dout$ differ, as is the case in Fig.~\ref{fig:explicit_setups}(a), we first construct the swap with the same number $\max(\din, \dout)$ of input and output ports. As the next step, we remove all redundant OAM exchangers and holo-beam splitters that correspond to the unused ports. From the symmetry of the swap, one can remove the elements from either of the two sides of the setup. Nevertheless, removing the exchangers from the output side is more resource-efficient than removing holo-beam splitters from the input side. Therefore, throughout the paper we assume that the number of output ports never exceeds the number of input ports. When necessary, the structure is used backward. In front of and after the main body of the swap operator, there is an additional path permutation. It turns out that the same permutations appear also in OAM sorters and path-only Fourier transforms and so in the resulting setup these path interconnections cancel each other out.

In Fig.~\ref{fig:explicit_setups} explicit setups for the $d$-dimensional OAM Fourier transform are shown for $d = 2$, $4$, and $8$. The setup for $d = 2$ is different in structure from the other powers of 2 and consists only of two exchangers and one 2-dimensional path-only Fourier transform, i.e., a beam splitter. The setup for $d = 4$ is the simplest setup for dimensions of the form $d = 2^{2K}$, for which the number of paths in the setup stays constant. Analogously, the setup for $d = 8$ is the simplest setup for dimensions of the form $d = 2^{2K+1}$, where the number of paths varies in different stages of the setup.

\end{document}